\documentclass{IEEEtran}
\pdfoutput=1
\usepackage{amsmath, amssymb, bm, cite, epsfig, psfrag}
\usepackage{multirow}
\usepackage{subfig}
\usepackage{graphicx}
\usepackage[font=small]{caption}
\def\beq{\begin{equation}}
\def\eeq{\end{equation}}
\def\beqa{\begin{eqnarray}}
\def\eeqa{\end{eqnarray}}
\def\beqan{\begin{eqnarray*}}
\def\eeqan{\end{eqnarray*}}

\def\span{\mathop{\mathrm{span}}}

\title{What Will 5G Be?}

\author{Jeffrey~G.~Andrews,~\IEEEmembership{Fellow,~IEEE,}
        Stefano~Buzzi,~\IEEEmembership{Senior Member,~IEEE,}
        Wan~Choi,~\IEEEmembership{Senior Member,~IEEE,}
        Stephen~Hanly,~\IEEEmembership{Member,~IEEE,}
        Angel~Lozano,~\IEEEmembership{Fellow,~IEEE,}
        Anthony~C.K.~Soong,~\IEEEmembership{Fellow,~IEEE,}
        Jianzhong~Charlie~Zhang,~\IEEEmembership{Senior Member,~IEEE}% <-this % stops a space
\thanks{J. G. Andrews (jandrews@ece.utexas.edu) is with the University of Texas at Austin, USA.}% <-this % stops a space
\thanks{S. Buzzi (buzzi@unicas.it) is with University of Cassino and Southern Lazio, Italy, and with CNIT, Italy. }% <-this % stops a space
\thanks{W. Choi (wchoi@kaist.edu) is with Korea Advanced Institute of Science and Technology (KAIST), Daejeon, South Korea.}% <-this % stops a space
\thanks{S. Hanly (stephen.hanly@mq.edu.au) is with Macquarie University, Sydney, Australia.}% <-this % stops a space
\thanks{A. Lozano (angel.lozano@upf.edu) is with Universitat Pompeu Fabra (UPF), Barcelona, Spain.}% <-this % stops a space
\thanks{A. C. K. Soong (anthony.soong@huawei.com) is with Huawei Technologies, Plano, Texas, USA.}% <-this % stops a space
\thanks{J. C. Zhang (jianzhong.z@samsung.com) is with Samsung Electronics, Richardson, Texas, USA.}% <-this % stops a space
\thanks{ Article last revised: \today}}

\begin{document}

% The paper headers
\markboth{IEEE JSAC special issue on 5G Wireless Communication Systems}%
{IEEE JSAC special issue on 5G Wireless Communication Systems}
\maketitle

\setcounter{page}{1}

\maketitle
\begin{abstract}
What will 5G be?  What it will \emph{not} be is an incremental advance on 4G. The previous four generations of cellular technology have each been a major paradigm shift that has broken backwards compatibility.  And indeed, 5G will need to be a paradigm shift that includes very high carrier frequencies with massive bandwidths, extreme base station and device densities and unprecedented numbers of antennas.  But unlike the previous four generations, it will also be highly integrative: tying any new 5G air interface and spectrum together with LTE and WiFi to provide universal high-rate coverage and a seamless user experience.  To support this, the core network will also have to reach unprecedented levels of flexibility and intelligence, spectrum regulation will need to be rethought and improved, and energy and cost efficiencies will become even more critical considerations.  This paper discusses all of these topics, identifying key challenges for future research and preliminary 5G standardization activities, while providing a comprehensive overview of the current literature, and in particular of the papers appearing in this special issue.
\end{abstract}

%********************************************************************************
% Introductory Section (JA)
%********************************************************************************

\section{Introduction}
\label{sec:intro}

\subsection{The Road to 5G}

In just the past year, preliminary interest and discussions about a possible 5G standard have evolved into a full-fledged conversation that has captured the attention and imagination of researchers and engineers around the world.  As the long-term evolution (LTE) system embodying 4G has now been deployed and is reaching maturity, where only incremental improvements and small amounts of new spectrum can be expected, it is natural for researchers to ponder ``what's next?'' \cite{bruno20093gpp}.  However, this is not a mere intellectual exercise.  Thanks largely the annual visual network index (VNI) reports released by Cisco, we have quantitative evidence that the wireless data explosion is real and will continue.  Driven largely by smartphones, tablets, and video streaming, the most recent (Feb. 2014) VNI report \cite{CiscoVNI14} and forecast makes plain that an incremental approach will not come close to meeting the demands that networks will face %come
by 2020.

In just a decade, the amount of IP data handled by wireless networks
will have increased by well over a factor of 100: from under 3
exabytes in 2010 to over 190 exabytes by 2018, on pace to exceed 500
exabytes by 2020.  This deluge of data has been driven chiefly by video thus far,
but new unforeseen applications can reasonably be expected to
materialize by 2020.  In addition to the sheer volume of data, the
number of devices and the data rates will continue to grow
exponentially.  The number of devices could reach the tens or even
hundreds of billions by the time 5G comes to fruition, due to many
new applications beyond personal communications
\cite{CLLPRT10,maeder2011challenge,hilton2010machine}. It is our
duty as engineers to meet these intense demands via innovative new
technologies that are smart and efficient yet grounded in reality.
Academia is engaging in large collaborative projects such as METIS \cite{METIS} and 5GNOW
\cite{5GNOW}, while the industry is driving
preliminary 5G standardization activities (cf. Sec.
\ref{sec:standards}).   To further strengthen these activities, the
public-private partnership for 5G infrastructure recently constituted in Europe
will funnel massive amounts of funds into related research
\cite{5GPPP}.

This article is an attempt to summarize and overview many of these exciting developments, including the papers in this special issue.  In addition to the highly visible demand for ever more network capacity, there are a number of other factors that make 5G interesting, including the potentially disruptive move to millimeter wave (mmWave) spectrum, new market-driven ways of allocating and re-allocating bandwidth, a major ongoing virtualization in the core network that might progressively spread to the edges, the possibility of an ``Internet of Things'' comprised of billions of miscellaneous devices, and the increasing integration of past and current cellular and WiFi standards to provide an ubiquitous high-rate, low-latency experience for network users.

This editorial commences with our view of the ``big three'' 5G technologies: ultra-densification, mmWave, and massive multiple-input multiple-output (MIMO). Then, we consider important issues concerning the basic transmission waveform, the increasing virtualization of the network infrastructure, and the need for greatly increased energy efficiency. Finally, we provide a comprehensive discussion of the equally important regulatory and standardization issues that will need to be addressed for 5G, with a particular focus on needed innovation in spectrum regulation.

\subsection{Engineering Requirements for 5G}
\label{sec:requirements}

In order to more concretely understand the engineering challenges facing 5G, and to plan to meet them, it is necessary to first identify the requirements for a 5G system.  The following items are requirements in each key dimension, but it should be stressed that not all of these need to be satisfied simultaneously.  Different applications will place different requirements on the performance, and peak requirements that will need to be satisfied in certain configurations are mentioned below. For example, very-high-rate applications such as streaming high-definition video may have relaxed latency and reliability requirements compared to driverless cars or public safety applications, where latency and reliability are paramount but lower data rates can be tolerated.

\subsubsection{Data Rate}  The need to support the %aforementioned
mobile data traffic explosion is unquestionably the main driver behind 5G.  %Throughput can be thought of or described in several different ways. The first is the \textbf{aggregate throughput} of the network, which corresponds to the total amount of data it can serve.  Since aggregate throughput is over all users in a given network, it is most easily characterized in units of bits/sec/area.  The general consensus is that the aggregate throughput (equivalently, per unit area) will need to increase by roughly 1000x from 4G to 5G.
Data rate can be measured in several different ways, and there will be a 5G goal target for each such metric:

a) \textbf{Aggregate data rate} refers to the total amount of data the network can serve, characterized in units of bits/s/area.  The general consensus is that this quantity will need to increase by roughly 1000x from 4G to 5G.

b) \textbf{Edge rate, or 5\% rate,} is the worst data rate that a user can reasonably expect to receive when in range of the network, and so is an important metric and has a concrete engineering meaning.  Goals for the 5G edge rate range from 100 Mbps (easily enough to support high-definition streaming) to as much as 1 Gbps. %what Samsung is advocating.
%Even meeting 100 Mbps will be extraordinarily challenging for 95\% of the users, we believe, even with large technological advances.
Meeting 100 Mbps for 95\% of users will be extraordinarily challenging, even with major technological advances.  This requires about a 100x advance since current 4G systems have a typical 5\% rate of about 1 Mbps, although the precise number varies quite widely depending on the load, cell size, and other factors.

c) \textbf{Peak rate} is the best-case data rate that a user can hope to achieve under any conceivable network configuration. The peak rate is a marketing number, devoid of much meaning to engineers, but in any case it will likely be in the range of tens of Gbps.

Meeting the requirements in (a)-(b), which are about 1000x and 100x current 4G technology, respectively, are the main focus of this paper.

\subsubsection{Latency}  Current 4G roundtrip latencies are on the order of about 15 ms, and are based on the 1 ms subframe time with necessary overheads for resource allocation and access.  Although this latency is sufficient for most current services, anticipated 5G applications include two-way gaming, novel cloud-based technologies such as those that may be touch-screen activated (the ``tactile Internet'' \cite{Fet14}), and virtual and enhanced reality (e.g., Google glass or other wearable computing devices).  As a result, 5G will need to be able to support a roundtrip latency of about 1 ms, an order of magnitude faster than 4G. In addition to shrinking down the subframe structure, such severe latency constraints may have important implications on design choices at several layers of the protocol stack and the core network (cf. Sect. \ref{sec:design}).

\subsubsection{Energy and Cost}  As we move to 5G, costs and energy consumption will, ideally, decrease, but at least they should not increase on a per-link basis.  Since the per-link data rates being offered will be increasing by about 100x, this means that the Joules per bit and cost per bit will need to fall by at least 100x.  In this article, we do not address energy and  cost in a quantitative fashion, but we are intentionally advocating technological solutions that promise reasonable cost and power scaling.  For example, mmWave spectrum should be 10-100x cheaper per Hz than the 3G and 4G spectrum below 3 GHz. Similarly, small cells should be 10-100x cheaper and more power efficient than macrocells.  A major cost consideration for 5G, even more so than in 4G due to the new BS densities and increased bandwidth, is the backhaul from the network edges into the core. We address backhaul and other economic considerations in Section~\ref{sec-economics}.   As for energy efficiency, we address this more substantially in Section~\ref{sec:energy}.

\subsection{Device Types and Quantities.} 5G will need to be able to efficiently support a much larger and more diverse set of devices.  With the expected rise of machine-to-machine communication, a single macrocell may need to support 10,000 or more low-rate devices, along with its traditional high-rate mobile users.  This will require wholesale changes to the control plane and network management relative to 4G, whose overhead channels and state machines are not designed for such a diverse and large subscriber base.

%********************************************************************************
% Key Technologies Section
%********************************************************************************
\section{Key Technologies to Get to 1000x Data Rate}
\label{sec:1000x}

Of the requirements outlined in Sect. \ref{sec:requirements}, certainly the one that gets the most attention is the need for radically higher data rates across the board.  Our view is that the required 1000x will, for the most part, be achieved through combined gains in three categories:
\begin{itemize}
\item[a)] Extreme densification and offloading to improve the area spectral efficiency. Put differently,
more active nodes per unit area and Hz.
\item[b)] Increased bandwidth, primarily by moving towards and into mmWave spectrum but also by making better use of WiFi's unlicensed spectrum in the 5 GHz band. Altogether, more Hz.
\item[c)] Increased spectral efficiency, primarily through advances in MIMO, to support more bits/s/Hz per node.
\end{itemize}

The combination of more nodes per unit area and Hz, more Hz, and more bits/s/Hz per node, will compound into \emph{many} more bits/s per unit area.
Other ideas not in the above categories, e.g., interference management through BS cooperation \cite{SenErk03,LanTse04,FosKar06,venkatesan2007network,Cadambe08,MMK08,venkatesan2009wimax,huang2009increasing,PetersHeath,GesHan10,SuhHoTse10,WangWang11,SimeoneNOW12,LozHea13} may also contribute improvements, but the lion's share of the surge in capacity should come from ideas in the above categories. In the remainder of this section, these are distilled in some detail.

%=============================
% Small cells and offloading (JA)
%=============================
\subsection{Extreme Densification and Offloading}
\label{sec:offloading}

A straightforward but extremely effective way to increase the network capacity is %simply
to make the cells smaller. This approach has been %proven
demonstrated over several cellular generations \cite{ChaAnd08,DohHeaLoz:Is-the-PHY-layer-dead:11}. %Starting with the deployment of
The first such generation, in the early 1980s, %(with cell sizes on the order of hundreds of square kilometers), cell sizes have been incrementally shrunk,
had cell sizes on the order of hundreds of square kms. Since then, those sizes have been progressively shrinking and by now they are often fractions of a square km in urban areas. In Japan, for instance, the spacing between BSs can be as small as two hundred meters, giving a coverage area well under a tenth of a square km. %\textbf{ XXX Anthony or Charlie, please verify this}. %In addition, and more recently,
Networks are now rapidly evolving \cite{And13} to include nested small cells such as picocells (range under 100 meters) and femtocells (WiFi-like range) \cite{AndCla12}, as well as distributed antenna systems \cite{HeaDAS13} that are functionally similar to picocells from a capacity and coverage standpoint but have all their baseband processing at a central site and share cell IDs.

Cell shrinking has numerous benefits, the most important being the reuse of spectrum across a geographic area and the ensuing reduction in the number of users competing for resources at each BS. Contrary to widespread belief, as long as power-law pathloss models hold the signal-to-interference ratio (SIR) is preserved as the network densifies \cite{DhiGan12}.\footnote{The power-law pathloss model ceases to apply in the near field, very close to the transmitter \cite{RamGan13}.}
%Also, $\alpha$ itself can empirically be observed to be an increasing function of $d$ in most environments, although this does not appear to erode the SINR distribution since it adds attenuation to the interference relative to the closer serving BS---that is, the received power is a function $d^{-\alpha}$ of the transmit-receive distance $d$ with $\alpha > 2$---the signal-to-interference plus noise ratio (SINR) does not decrease as the network densifies \cite{DhiGan12}.
Thus, in principle, cells can shrunk almost indefinitely without a sacrifice in SIR, until nearly every BS serves a single user (or is idle). This allows each BS to devote its resources, as well as its backhaul connection, to an ever-smaller number of users.

As the densification becomes extreme, some challenges arise:
%The main challenges to reaping gains from extreme densification are:
\begin{itemize}
\item Preserving the expected cell-splitting gains as each BS becomes more lightly loaded, particularly low-power nodes.
\item Determining appropriate associations between users and BSs across multiple radio access technologies (RATs), which is crucial for optimizing the edge rate.
\item Supporting mobility through such a highly heterogeneous network.
\item Affording the rising costs of installation, maintenance and backhaul.
\end{itemize}
We next briefly discuss these challenges, particularly in view of the other technologies raised in this article.

\subsubsection{Base Station Densification Gains}

%We define the \emph{base station densification gain} $\rho > 0$ as the \emph{effective} increase in data rate relative to the increase in network density (a proxy here for cost).  Specifically, if we increase the overall BS density (in BSs/km$^2$ including all types of BSs) from $\lambda_1$  to $\lambda_2$, then
%\begin{equation}
%\rho = \frac{R_2 \lambda_1}{R_1 \lambda_2}
%\end{equation}
%where $R_1$ and $R_2$ are the before and after data rates (could be any measure thereof, e.g., edge or aggregate, which will naturally give different values of $\rho$.)
%For example, if the network density is doubled, and the aggregate data rate also doubles, then the densification gain is $\rho = 1$: the increase in BS density %had
%has an exactly proportional payoff in terms of achieved rates in this case.

We define the BS densification gain $\rho > 0$ as the \emph{effective} increase in data rate relative to the increase in network density, which is a proxy here for cost.  Specifically, if we achieve a data rate $R_1$ (could be any measure thereof, e.g., edge rate or aggregate) when the BS density is $\lambda_1$ BSs/km$^2$ and then we consider a higher BS density $\lambda_2$, with corresponding rate $R_2$, %increase the overall BS density (including all types of BSs) from $\lambda_1$  to $\lambda_2$ BSs/km$^2$, then
then the corresponding densification gain is
\begin{equation}
\rho = \frac{R_2 \lambda_1}{R_1 \lambda_2}.
\end{equation}
For example, if the network density is doubled, and the aggregate data rate also doubles, then the densification gain is $\rho = 1$: the increase in BS density %had
has an exactly proportional payoff in terms of achieved rates in this case.

In an interference-limited network with full buffers, the signal-to-interference-plus-noise ratio (SINR) is essentially equal to the SIR and, because the SIR distribution remains approximately constant as the network densifies, the best case scenario is $\rho \approx 1$.  In reality, buffers are not always full, and small cells tend to become more lightly loaded than macrocells as the network densifies.  Therefore, the SINR usually increases with density: in a noise-limited network by increasing the received signal power, and in interference-limited networks because the lightly loaded small cells generate less interference (while still providing an option for connectivity) \cite{DhiAnd13}.  Nevertheless, at microwave frequencies the gain in SINR is not enough to keep up with the decrease in small cell utilization, and thus $\rho < 1$.
 %Asymptotically as the small cell density increases, $\rho$ shrinks towards $0$ since the small cells begin to compete for users amongst themselves and become ever more lightly loaded, and the rate increase per cell cannot keep up with the density increase.
In an extreme case, consider $\lambda_1$ and $R_1$ held fixed with $\lambda_2 \rightarrow \infty$. In this asymptotic setting, the small cells compete for a finite pool of UEs, becoming ever more lightly loaded, and thus $\rho \rightarrow 0$.

Empirically and theoretically, we also observe that $\rho$ improves and can approach $1$ with macro-BS muting (eICIC in 3GPP) vs. the macrocells transmitting all the time and thus interfering with the small cells all the time.  This observation is relevant because the result is not obvious given that the macrocells are the network bottleneck.

An intriguing aspect of mmWave frequencies is that densification gains $\rho \gg 1$ may be possible. This is because, as discussed in Section \ref{sec-mmWave}, at these frequencies communication is largely noise-limited and increasing the density not only splits the cell resources and lightens the load, but it may increase the SINR dramatically.  As a %poignant
striking example of this, it was recently showed that, under a plausible urban grid-based deployment, increasing the BS count in a given area from 36 to 96---which decreased the inter-BS distance from 170 meters down to 85 meters---increased the $5\%$ cell-edge rate from 24.5 Mbps up to 1396 Mbps, giving \cite{NSN13}
\begin{equation}
\rho = \frac{1396 \cdot 36}{24.5 \cdot 96} = 21.3 .
\end{equation}
While conceding that this massive densification gain corresponds to a particular setup and model, it is nevertheless remarkable.

In general, quantifying and optimizing the densification gains in a wide variety of deployment scenarios and network models is a key area for continued small cell research.

\subsubsection{Multi-RAT Association}

Networks will continue to become increasingly heterogeneous as we move towards 5G.  A key feature therein will be increased integration between different RATs, with a typical 5G-enabled device having radios capable of supporting not only a potentially new 5G standard (e.g., at mmWave frequencies), but also 3G, numerous releases of 4G LTE including possibly LTE-Unlicensed \cite{QualcommLTEU}, several types of WiFi, and perhaps direct device-to-device (D2D) communication, all across a great many spectral bands. Hence, determining which standard(s) and spectrum to utilize and which BS(s) or users to associate with will be a truly complex task for the network \cite{Gali1409:Capturing}.

\begin{figure}
\centering
\includegraphics[width=3.5in]{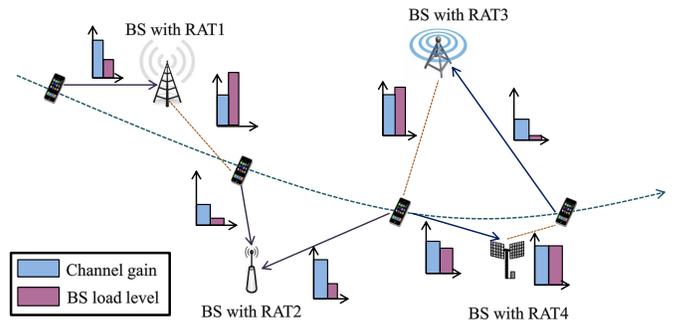}
    \caption{User association in a multi-RAT network over many frequency bands is complex. In this simplified scenario, a mobile user in turn associates with different BSs based on a tradeoff between the gain to that BS and the traffic load (congestion) that it is experiencing.  }
    \label{fig:MultiRAT}
\end{figure}

%Fundamentally,
Determining the optimal user association %is
can be a massive combinatorial optimization problem that depends on the SINR from every user to every BS, the instantaneous load at each BS, the choices of other users in the network, and possibly other constraints such as the requirement to utilize the same BS and standard in both uplink and downlink (to facilitate functioning control channels for resource allocation and feedback) \cite{YeRon13,CoFaMa12}.  Therefore, simplified procedures must be adopted \cite{And14}, an example of which appears in this special issue \cite{Shen1409:Distributed}.  Even a simple, seemingly highly suboptimal association approach based on aggressive but static biasing towards small cells and blanking about half of the macrocell transmissions
%(despite them being the network bottleneck)
has been shown to increase edge rates by as much as $500\%$ \cite{SinAnd13}.

The joint problem of user association and resource allocation in two-tier heterogeneous networks (HetNets), with adaptive tuning of the biasing and blanking in each cell, is considered in \cite{CoFaMa12,YeRon13,fooladivanda2013joint,bedekar2013optimal,BHWVTC13,BHWICC13,Deb2013,SorPed13}. An interesting model of hotspot traffic is considered in \cite{BHWVTC13,BHWICC13,bedekar2013optimal} where it is shown that, under various network utility metrics, the optimal cell association is determined by rate ratio bias, rather than power level bias.
%This was independently observed for a similar model in \cite{bedekar2013optimal}.

It will be interesting to extend these models to more general scenarios, including more than two tiers. A dynamic model of cell range expansion is considered in \cite{HW14}, where traffic arrives as a Poisson process in time and the feasible arrival rates, for which a stabilizing scheduling policy exists, are characterized. User association and load balancing in a HetNet, with massive MIMO at the BSs, is considered in \cite{BBPC14}. The problem of determining the optimal associations when there are multiple RATS, operating at different frequencies and using different protocols, has not yet received much attention. However, an interesting game theoretic approach is taken in \cite{AK-HWC13} to the RAT-selection problem, where convergence to Nash equilibria and the Pareto-efficiency of these equilibria are studied.  A related paper in this special issue \cite{Gao1409:Bargaining} explores the interaction between cellular operators and WiFi network owners.

%Adaptively tuning the biasing and blanking on a per-cell basis promises further gains \cite{BHWVTC13,BHWICC13,SorPed13}.
Adding mmWave into the picture adds significant additional complexity, since even the notion of a cell boundary is blurry at mmWave frequencies given the strong impact of blockages, which often result in nearby BSs being bypassed in favor of farther ones that are unblocked (cf. Fig. \ref{fig:mmWaveAssociations}).  On the positive side, interference is much less important in mmWave %, as we shall see, and
(cf. Section~\ref{sec-mmWave}) and thus the need for blanking %would be
is reduced.

In summary, there is a great deal of scope for modeling, analyzing and optimizing BS-user associations in 5G.

\subsubsection{Mobility Support}

Clearly, the continued network densification and increased heterogeneity poses challenges for the support of mobility.
Although a hefty share of data is served to stationary indoor users, the support of mobility and always-on connectivity is arguably the single most important feature of cellular networks relative to WiFi.
Because modeling and analyzing the effect of mobility on network performance is difficult, we expect to see somewhat ad hoc solutions such as in LTE Rel-11 \cite{lee2012coordinated} where user-specific virtual cells are defined to distinguish the physical cell from a broader area where the user can roam without the need for handoff, communicating with any BS or subset of BSs in that area. Or in mmWave, restricting highly mobile users to macrocells and microwave frequencies, thereby forcing them to tolerate lower rates. 
Handoffs will be particularly challenging at mmWave frequencies since transmit and receive beams must be aligned to communicate.
Indeed, the entire paradigm of a handoff initiated and managed at layer 3 by the core network will likely not exist in 5G; instead, handoffs may be opportunistic, based on mmWave beam alignments, or indistinguishable from PHY/MAC interference management techniques whereby users communicate with multiple coordinated BSs, as exemplified by
%(multi-point connectivity, i.e. CoMP), %e.g. \cite{Kim1409:Virtual}.
\cite{Kim1409:Virtual} in this special issue. 

\subsubsection{Cost}

Evolving to ever-smaller cells requires ever-smaller, lower-power and cheaper BSs, and %asymptotically
there is no fundamental reason a BS needs to be more expensive than a user device or a WiFi node \cite{And13}.  Nevertheless, obtaining permits, ensuring fast and reliable backhaul connections, and paying large monthly site rental fees for operator-controlled small-cell placements have proven a major hindrance to the growth of picocell, distributed antennas, and other enterprise-quality small cell deployments.  Of these, only the backhaul is primarily a technical challenge.  Regulatory reforms and infrastructure sharing (cf. Section~\ref{sec-economics}) may help address the other challenges.

Turning to end-user-deployed femtocells and WiFi access points, these are certainly much more cost-effective both from a capital and operating expense perspective \cite{ChaAnd08}.  However, major concerns exist here too.  These include the coordination and management of the network to provide enterprise-grade service, which given the scale of the deployments requires automated self-organization \cite{SONSurvey13}.   A further challenge is that these end-user deployments utilize the end-user's backhaul connection and access point, both of which the end-user has a vested interest in not sharing, and in some countries a legal requirement not to.  Anecdotally, all readers of this article are familiar with the scenario where a dozen WiFi access points are within range, but all are secured and inaccessible.  From an engineering perspective, this closed-access status quo is highly inefficient and the cost for 5G would be greatly reduced in an open-access paradigm for small cells.  One preliminary but successful example is Fon, which as of press time boasts over 13 million shared WiFi access points.\\

5G and all networks beyond it will be extremely dense and heterogeneous, which introduces many new challenges for network modeling, analysis, design and optimization.  We further discuss some of the nonobvious intersections of extreme densification with mmWave and massive MIMO, respectively, in the next two sections.
Before proceeding, however, we briefly mention that besides cell shrinking a second approach to densification exists in the form of direct D2D communication.
This allows users in close proximity to establish direct communication, replacing two long hops via the BS with a single shorter hop. 
Provided there is sufficient spatial locality in the wireless traffic, this can bring about reduced power consumption and/or higher data rates, and a diminished latency \cite{Kaufman-D2D2008, Hao-Doppler-2013,WirelessMagDec2010}.
Reference \cite{Nade1409:ITLinQ} in this special issue proposes a novel way of scheduling concurrent D2D transmissions so as to densify while offering interference protection guarantees.

\begin{figure}
\centering
\includegraphics[width=3.5in]{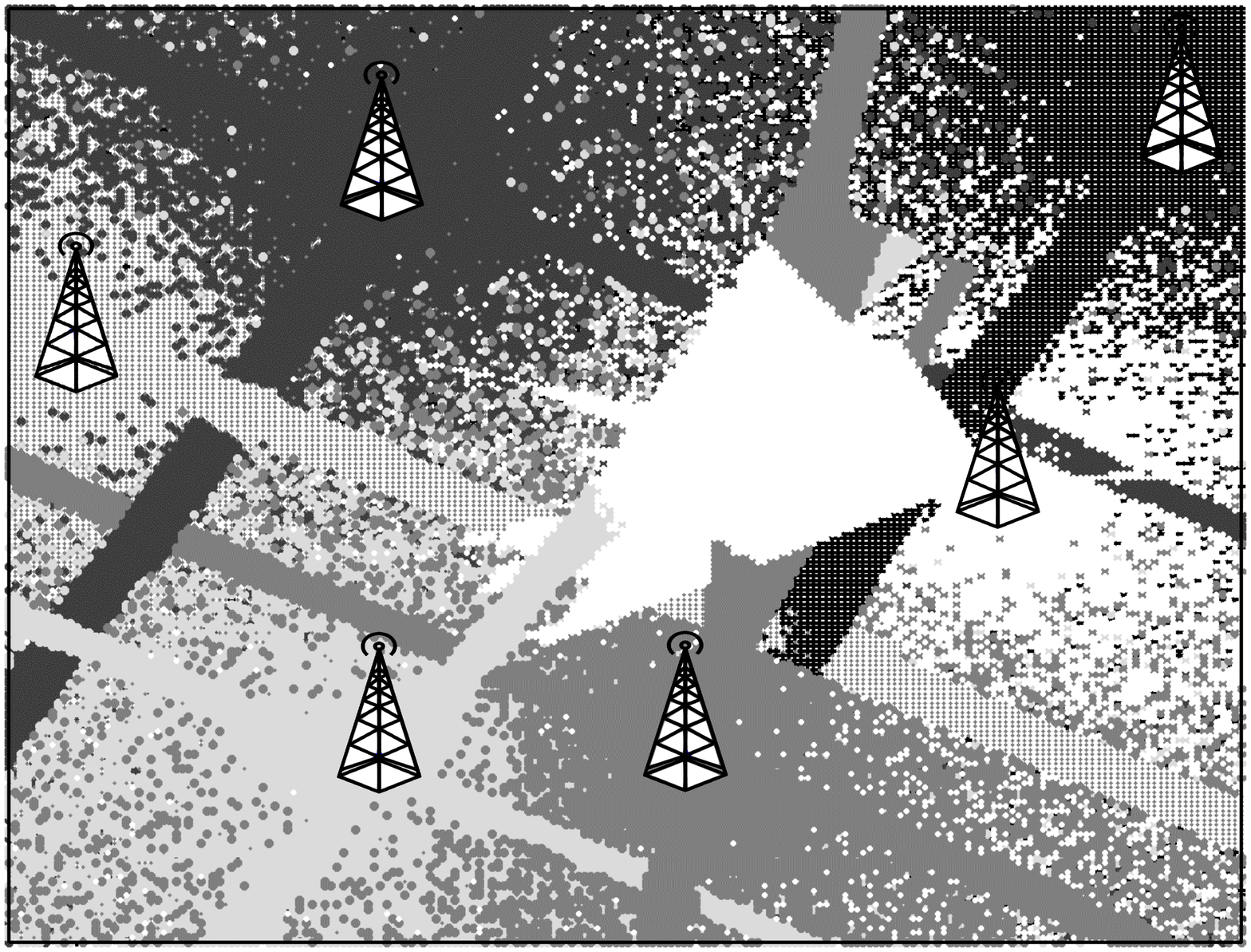}
\includegraphics[width=3.5in]{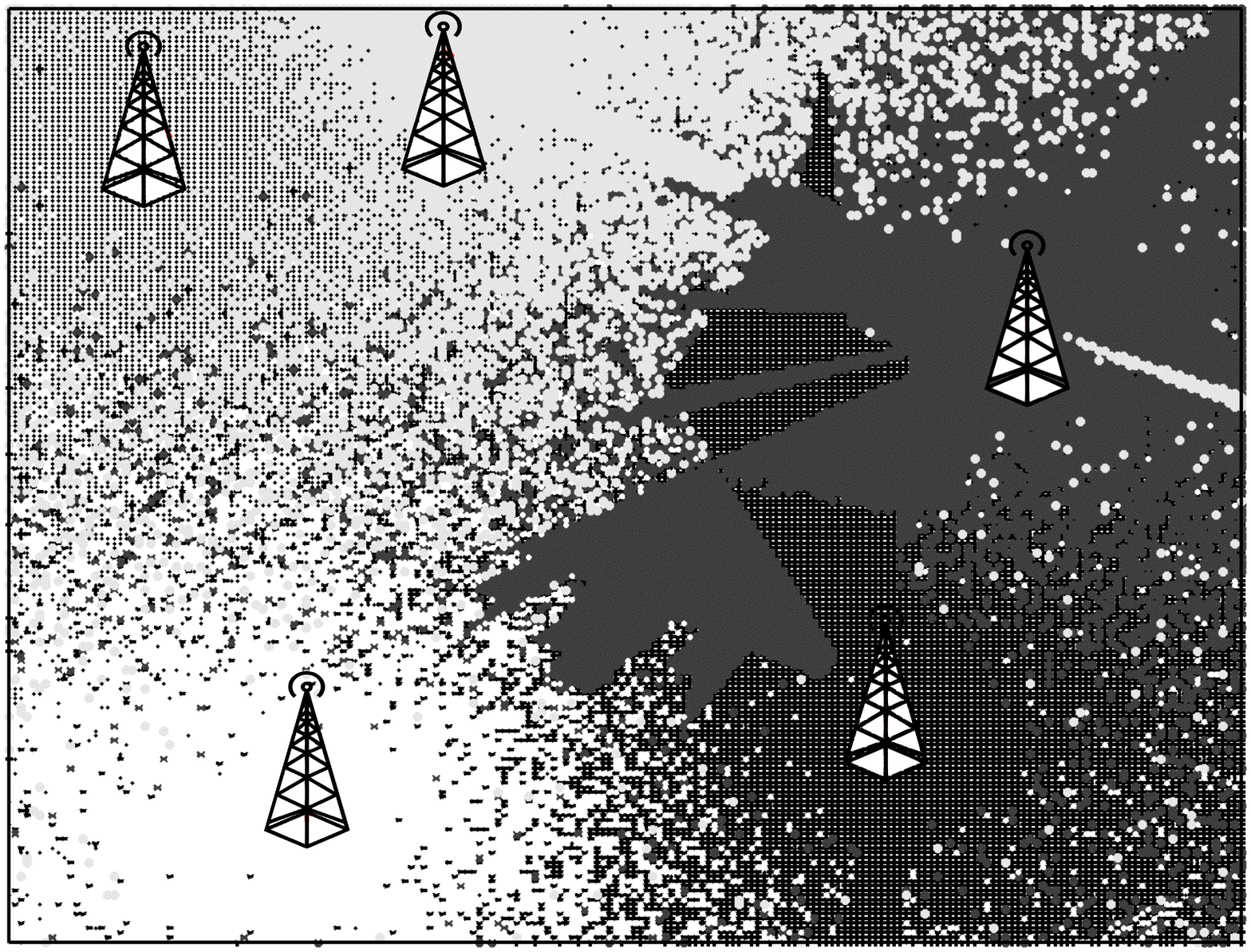}
    \caption{Calculated mmWave BS associations with real building locations \cite{SinKul14}.  The shaded regions correspond to association with the BS centered at that shade.  Blocking, LOS vs. non-LOS propagation, and beam directionality render our usual notion of cell boundaries obsolete. }
    \label{fig:mmWaveAssociations}
\end{figure}

%There is also no substitute for continued careful network planning whereby small cells are placed near traffic bottlenecks, although presumably much of this planning can be automated given suitable data collection.

%To do: (1) integrate references especially from this S.I.. and others' recent work; (2) make some of the discussion more rigorous and (3) verify the various statements (Anthony and Charlie). (4) add a figure showing multi-RAT or mmWave associations.  

%=============================
% Millimeter Wave (WC)
%=============================
\subsection{Millimeter Wave}

\label{sec-mmWave}

Terrestrial wireless communication systems have largely restricted their operation to the relatively slim range of microwave frequencies that extends from several hundred MHz to a few GHz and corresponds to wavelengths in the range of a few centimeters up to about a meter. %Within this range, wave propagation is robust and resonant antennas occupy a sweet spot: sufficiently small to be used on portable devices, but sufficiently large to radiate and capture a substantial amount of electromagnetic energy.  
By now though, this spectral band---often called ``beachfront spectrum''---has become nearly fully occupied, in particular at peak times and in peak markets.  Regardless of the efficacy of densification and offloading,  much more bandwidth is needed \cite{BroadbandPlan,ITU2078}.

Although beachfront bandwidth allocations can be made significantly more efficient by modernizing regulatory and allocation procedures, as discussed in Section \ref{sec-spectrum}, to put large amounts of new bandwidth into play there is only one way to go: up in frequency. Fortunately, vast amounts of relatively idle spectrum do exist in the mmWave range of 30--300 GHz, where wavelengths are 1--10 mm.
There are also several GHz of plausible spectrum in the 20--30 GHz range.

The main reason that mmWave spectrum lies idle is that, until recently, it had been deemed unsuitable for mobile communications because of rather hostile propagation qualities, including strong pathloss, atmospheric and rain absorption, low diffraction around obstacles and penetration through objects, and, further, because of strong phase noise and exorbitant equipment costs.  The dominant perception had therefore been that such frequencies, and in particular the large unlicensed band around 60 GHz \cite{4457895}, were suitable mainly for very-short-range transmission \cite{Vaughan-Nicholos2010, BaykasSumLanWang2011,RappaportMurdockGutierrez2011}. Thus, the focus had been on WiFi (with the WiGiG standard in the 60-GHz band) and also
on fixed-wireless applications in the 28, 38, 71--76 and 81--86 GHz. However, semiconductors are maturing, their costs and power consumption rapidly falling---largely thanks to the progress of the aforementioned short-range standards---and the other obstacles related to propagation are now considered increasingly surmountable given time and focused effort  \cite{MarcusPattan2005,AlejosSanchezCuinas2008, PiKhan2011, RappaportEtAl2013, RohMag14, RanRapErk14}.

% Took out these from above citation, too many from same group, and these are conference papers.   Also removed below.  AzarEtAl2013ICC,ZhaoEtAl2013ICC,Samimi2013VTC,

\subsubsection{Propagation Issues}
Concerning mmWave propagation for 5G cellular communication, the main issues under investigation are:

\textbf{Pathloss}. If the electrical size of the antennas (i.e., their size measured by the wavelength $\lambda=c/f_{\rm c}$ where $f_{\rm c}$ is the carrier frequency) is kept constant, as the frequency increases the antennas shrink and their effective aperture scales with $\frac{\lambda^2}{4 \pi}$; then, 
the free-space pathloss between a transmit and a receive antenna grows with $f^2_{\rm c}$.
%due to the effective antenna aperture scaling with $\frac{\lambda^2}{4 \pi}$ where $\lambda=c/f_{\rm c}$.
Thus, increasing $f_{\rm c}$ by an order of magnitude, say from 3 to 30 GHz, adds 20 dB of power loss regardless of the transmit-receive distance.   However, if the antenna aperture at one end of the link is kept constant as the frequency increases, then the free-space pathloss remains unchanged. Further, if both the transmit and receive antenna apertures are held constant, then the free-space pathloss actually \emph{diminishes} with $f^2_{\rm c}$: a power gain that would help counter the higher noise floor associated with broader signal bandwidths.

Although preserving the electrical size of the antennas is desirable for a number of reasons,
maintaining at the same time the aperture %entails utilizing
is possible utilizing arrays, which aggregate the individual antenna apertures: %The shrinking of each individual antenna makes this feasible, enabling large arrays in a small area.
as the antennas shrink with frequency, progressively more of them must be added in the original area.
The main challenge becomes cophasing these antennas so that they steer and/or collect energy productively.  This challenge becomes more pronounced when the channel changes rapidly, for example due to mobility and the higher Doppler shifts at mmWave frequencies or due to rapid alterations in the physical orientation of the devices. 

\textbf{Blocking}.
MmWave signals exhibit reduced diffraction and a more specular propagation than their microwave counterparts, and hence they are much more susceptible to blockages.   This results in a nearly bimodal channel depending on the presence or absence of Line-of-Sight (LoS).
According to recent measurements \cite{RappaportEtAl2013,RanRapErk14}, as the transmit-receive distance grows the pathloss accrues close to the free-space value of 20 dB/decade under LoS propagation, but drops to 40 dB/decade plus an additional blocking loss of 15--40 dB otherwise.  Because of the sensitivity to blockages, a given link can rapidly transition from usable to unusable and, unlike small-scale fading, large-scale obstructions cannot be circumvented with standard small-scale diversity countermeasures.   New channel models capturing these effects are much needed, and in fact currently being developed \cite{RappaportGutierrezBen-DorMurdockQiaoTamir2013,
RappaportEtAl2013,KulSin14} and applied to system-level analysis \cite{Bai2013,BaiHeath13a,Bai2012,SinKul14} and simulation studies such as \cite{ghoshmillimeter} and \cite{AkdenizaLiuSunRanganRappaportErkip2014} in this special issue.

\textbf{Atmospheric and rain absorption}.
The absorption due to air and rain is noticeable, especially the 15 dB/km oxygen absorption within the 60-GHz band (which is in fact why this band is unlicensed), but it is inconsequential for the urban cellular deployments currently envisioned \cite{MarcusPattan2005, PiKhan2011} where BS spacings might be on the order of 200 m.   In fact, such absorption is beneficial since it further attenuates background interference from more distant BSs, effectively increasing the isolation of each cell.

The main conclusion is that the propagation losses for mmWave frequencies are surmountable, but require large antenna arrays to steer the beam energy and collect it coherently.  While physically feasible, the notion of narrow-beam communication is new to cellular communications and poses difficulties, which we next discuss.

\subsubsection{Large arrays, narrow beams}

Building a cellular system out of narrow and focused beams is highly nontrivial and changes many traditional aspects of cellular system design.  MmWave beams are highly directional, almost like flashlights, which completely changes the interference behavior as well as the sensitivity to misaligned beams.  The interference adopts an on/off behavior where most beams do not interfere, but %then sporadically strong interference may materialize very suddenly.  
strong interference does occur intermittently. Overall, interference is de-emphasized and mmWave cellular links may often be noise-limited, which is a major reversal from 4G. Indeed, even the notion of a ``cell'' is likely to be very different in a mmWave system since, rather than distance, blocking is often the first-order effect on the received signal power. This is illustrated in Fig.~\ref{fig:mmWaveAssociations}.

\textbf{Link acquisition.} A key challenge for narrow beams is the difficulty in establishing associations between users and BSs, both for initial access and for handoff. To find each other, a user and a BS may need to scan lots of angular positions where a narrow beam could possibly be found, or deploy extremely large coding/spreading gains over a wider beam that is successively narrowed in a multistage acquisition procedure.  Developing solutions to this problem, particularly in the context of high mobility, is an important research challenge.

\begin{figure}[!t]
\centering
\includegraphics[width=3.5in]{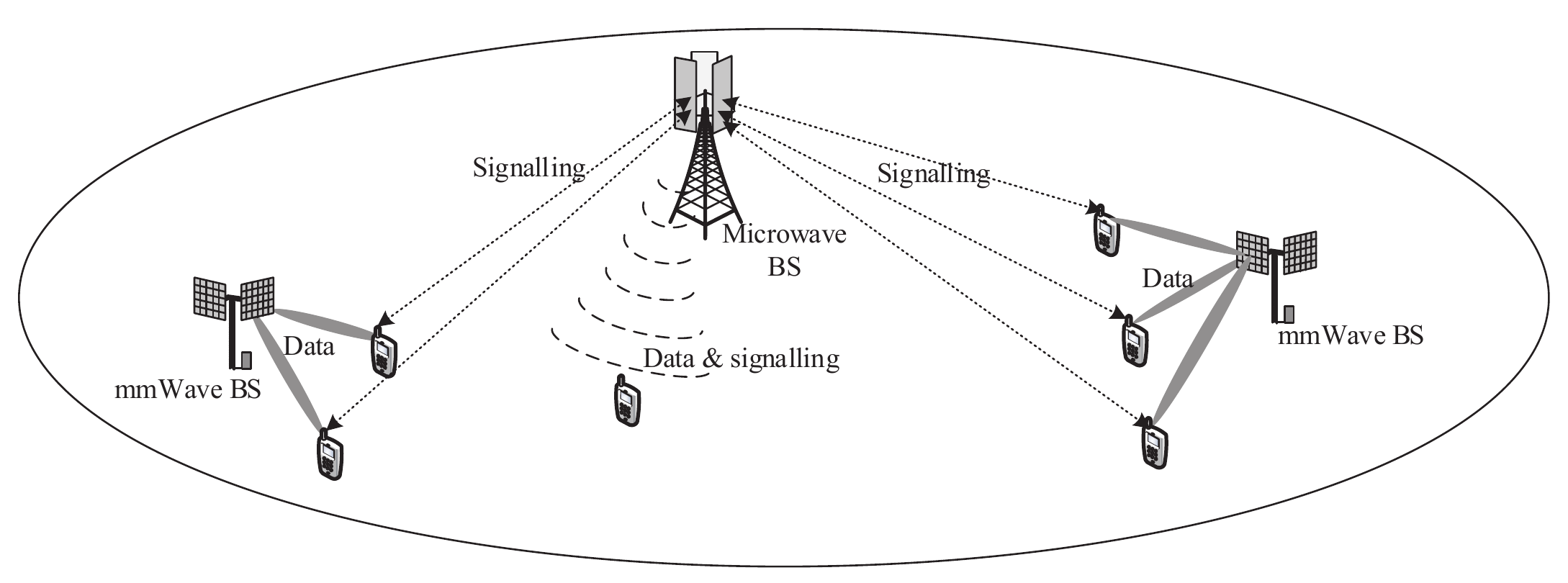}
   % \centerline{\psfig{figure=figs/architecture.eps,width=0.8\columnwidth} }
    \caption{MmWave-enabled network with phantom cells.}
    \label{Fig:Architecture}
\end{figure}

\textbf{Leveraging the legacy 4G network.}  A concurrent utilization of microwave and mmWave frequencies could
go a long way towards overcoming some of the above hurdles. An interesting proposal in that respect is the notion of
``phantom cells'' (relabeled ``soft cells'' within 3GPP) \cite{IshiiKishiyamaTakahashi2012Globecom},
where mmWave frequencies would be employed for payload data transmission from small-cell BSs
while the control plane would operate at microwave frequencies from macro BSs (cf. Fig. \ref{Fig:Architecture}). This would ensure stable and reliable control connections, based on which blazing fast data transmissions could be arranged over short-range mmWave links \cite{BaiHea14-Asilomar}. Sporadic interruptions of these mmWave links would then be far less consequential, as control links would remain in place and lost data could be recovered through retransmissions.

\textbf{Novel transceiver architectures needed.}  Despite the progress made in WiFi mmWave systems, nontrivial hardware issues remain, and in some cases will directly affect how the communication aspects are designed.  Chief among these is the still-exorbitant power consumption of particularly the analog-to-digital (A/D) but also the digital-to-analog (D/A) converters needed for large bandwidths.  A main consequence is that, although large antenna arrays and high receiver sensitivities are needed to deal with the pathloss, having customary fully digital beamformers for each antenna
appears to be unfeasible. More likely are structures based on old-fashioned analog phase shifters or, perhaps,
hybrid structures where groups of antennas share a single A/D and D/A  \cite{ZhangMolishKung2005, WangEtal2009, VenkateswaranVeen2010, HurKimLoveKrogmeierThomasGhosh2013}. On the flip side, offering some relief from these difficulties, the channels are sparser and thus the acquisition of channel-state information is facilitated; in particular, channel estimation and beamforming techniques exploiting sparsity in the framework of compressed sensing are being explored \cite{AyachRajagopaAbu-SurraPiHeath2014, AlkhateebAyachLeusHeath2014}.

%=============================
% Massive MIMO subsection (AL)
%=============================
\subsection{Massive MIMO}
\label{sec:MIMO}

Stemming from research that blossomed in the late 1990s \cite{FosGan98,Tel99}, MIMO communication was introduced into WiFi systems around 2006 and into 3G cellular shortly thereafter.
In essence, MIMO embodies the spatial dimension of the communication that arises once a multiplicity of antennas are available at base stations and mobile units.
If the entries of the channel matrix that ensues exhibit---by virtue of spacing, cross-polarization and/or angular disposition---sufficient statistical independence, multiple spatial dimensions become available for signaling and the spectral efficiency multiplies accordingly \cite{farrokhi2002spectral,lozano2002capacity}.

In single-user MIMO (SU-MIMO), the dimensions are limited by the number of antennas that can be accommodated on a portable device. However, by having each BS communicate with several users concurrently, the multiuser version of MIMO (MU-MIMO) can
effectively pull together the antennas at those users and overcome this bottleneck.
Then, the signaling dimensions are given by the smallest between the aggregate number of antennas at those users and the number of antennas at the BS.

Furthermore, in what is now known as coordinated multipoint (CoMP) transmission/reception, multiple BSs
can cooperate and act as a single effective MIMO transceiver thereby turning some of the interference in the system into useful signals; this concept in fact underpins many of the approaches to interference and mobility management mentioned earlier in this section.

Well-established by the time LTE was developed, MIMO was a native ingredient thereof with two-to-four antennas per mobile unit and as many as eight per base station sector, and it appeared that, because of form factors and other apparent limitations, such was the extent to which MIMO could be leveraged. Marzetta was instrumental in articulating a vision in which the number of antennas increased by more than an order of magnitude, first in a 2007 presentation \cite{Mar07} with  the details formalized in a landmark paper \cite{Marzetta2010}. The proposal was to equip BSs with a number of antennas much larger than the number of active users per time-frequency signaling resource, and given that under reasonable time-frequency selectivities accurate channel estimation can be conducted for at most some tens of users per resource, this condition puts the number of antennas per base station into the hundreds. This bold idea, initially termed ``large-scale antenna systems'' but now more popularly known as ``massive MIMO'', offers enticing benefits:
\begin{itemize}
\item Enormous enhancements in spectral efficiency without the need for increased BS densification, with the possibility---as is always the case---of trading some of those enhancements off for power efficiency improvements \cite{HoydisBrinkDebbah2013,RusekPerssonLauLarssonMarzettaEdforsTufvesson2013}.
\item Smoothed out channel responses because of the vast spatial diversity, which brings about the favorable action of the law of large numbers. In essence, all small-scale randomness abates as the number of channel observations grows.
\item Simple transmit/receive structures because of the quasi-orthogonal nature of the channels between each BS and the set of active users sharing the same signaling resource. For a given number of active users, such orthogonality sharpens as the number of BS antennas grows and simple linear transceivers, even plain single-user beamforming, perform close-to-optimally.
\end{itemize}

The promise of these benefits has elevated massive MIMO to a central position in preliminary discussions about 5G \cite{BoccardiLozanoMarzettaPopovski2014Disruptive}, with a foreseen role of providing a high-capacity umbrella of ubiquitous coverage in support of underlying tiers of small cells. However, for massive MIMO to become a reality, several major challenges must first be overcome, and the remainder of this section is devoted to their dissection. For very recent contributions on these and other aspects, the reader is referred to a companion special issue on massive MIMO \cite{matthaiou2013guest}.
The present special issue contains further new contributions, mentioned throughout the discussion that follows, plus reference \cite{Xian1409:Massive} dealing with the massification of MIMO multicasting \cite{sidiropoulos2006transmit,lozano2007long}.

%\subsubsection{Pilot Contamination}
\subsubsection{Pilot Contamination and Overhead Reduction}

Pilot transmissions can be made orthogonal among same-cell users, to facilitate cleaner channel estimates \cite{HasHoc03,jindal2010unified}, but must be reused across cells---for otherwise all available resources would end up consumed by pilots. This inevitably causes interference among pilots in different cells and hence puts a floor on the quality of the channel estimates. This interference, so-called ``pilot contamination,'' does not vanish %with the number of base antennas
as the number of BS antennas grows large, and so is the one impairment that remains asymptotically. However, pilot contamination is a relatively secondary factor for all but colossal numbers of antennas \cite{HUH2012}. Furthermore, various methods to reduce and even eliminate pilot contamination via low-intensity BS coordination have already been formulated \cite{ASHIKHMIN12,YIN2013}. Still, a careful design of the pilot structures is required to avoid an explosion in overhead. The ideas being considered to reign in pilot overheads include exploiting spatial correlations, so as to share pilot symbols among antennas, and also segregating the pilots into classes (e.g., channel strength gauging for link adaptation v. data detection) such that each class can be transmitted at the necessary rate, and no faster.
% those pilots whose quantity scales with the number of antennas can be transmitted at a lower rate.

\subsubsection{Architectural Challenges}

A more serious challenge to the realization of the massive MIMO vision has to do with its architecture. The vision requires radically different BS structures where, in lieu of a few high-power amplifiers feeding a handful of sector antennas, we would have a myriad of tiny antennas fed by correspondingly low-power amplifiers; most likely each antenna would have to be integrated with its own amplifier. Scalability, antenna correlations and mutual couplings, and cost, are some of the issues that must be sorted out.
At the same time, opportunities arise for innovative topologies such as conformal arrays along rooftops or on building facades, and we next dwell on a specific topological aspect in which innovation is taking place.

Within this special issue, \cite{Zeng1409:Electromagnetic} explores alternative and highly innovative antenna designs based on the utilization of an electromagnetic lens-focusing antenna.
 %to reduce the signal processing and hardware cost.
% and that are touched upon by \cite{Zeng1409:Electromagnetic} in this special issue.

%...and the term ``full-dimension MIMO (FD-MIMO)'' has been coined to refer to 2D arrays \cite{NAM20}.

\subsubsection{Full-Dimension MIMO and Elevation Beamforming}

%The recent popularity of various smart phone multimedia applications led to dramatic increase of wireless data traffic in carriers’ networks across the world.
%In the current  3GPP LTE and LTE-Advanced standards (Release 8 -11) , several new features were introduced to improve the cell capacity as well as the cell edge throughput of the wireless network, including as multiple-input multiple-output (MIMO) with up to 8 antenna ports, coordinated multipoint (CoMP) transmission and reception, and heterogeneous networks.

Existing BSs mostly feature linear horizontal arrays, which in tower structures can only accommodate %so many
a limited number of antennas, due to form factors, and which only exploit the azimuth angle dimension. %for beamforming and does not fully exploit all spatial dimensions.
By adopting planar 2D arrays similar to  Fig.~\ref{Fig:Architecture} and further exploiting the elevation angle, so-called full-dimension MIMO (FD-MIMO) can house many more antennas with the same form factor \cite{Nam2013}. As a side benefit, tailored vertical beams increase the signal power and reduce interference to users in neighboring cells.  Some preliminary cell average and edge data rates obtained from Samsung's network simulator are listed in Table~\ref{tab:FD-MIMO-Tput}
where, with numbers of antennas still modest for what massive MIMO is envisioned to be, multiple-fold improvements are observed.
%showing a capacity gain of up to 3 times and 6 times, respectively.
%While these results are very promising, there is room for further improvement as 2D-array-specific precoding and scheduling algorithm is still an open topic under active investigation.

%\begin{figure}
%\centering \includegraphics[width=4.5in]{figs/FD-MIMO-system.pdf}
%\caption{FD-MIMO BS.}
%\label{fig:FD-MIMO-system}
%\end{figure}

%The introduction of the elevation dimension improves user SINR by pointing the vertical beam pattern toward the direction of the user of interest. This increases the received signal power and reduces the interference experiences by users in neighbouring sectors. In an FD-MIMO system with a 2D active array, UE-specific elevation beamforming is combined with UE-specific azimuth beamforming to support high-order MU-MIMO.
%From 3GPP standards perspective, the 3D channel modelling study is expected to complete in 2014.
%and the focus will be then shifted toward system design aspects such as reference signal design, codebook and feedback mechanism for TDD and FDD systems, control signalling support for high order MU-MIMO and necessary RF requirements.

%\begin{figure}
%\centering \includegraphics[width=3.5in]{figs/FD-MIMO-Tput.pdf}
%\caption{\small FD-MIMO system-level downlink simulation results. Half-wavelength antenna spacing in both the horizontal and vertical dimensions at the base stations, 2 antennas per user. The baseline is SU-MIMO with 4 antennas per BS and the FD-MIMO results are for MU-MIMO with 16 and 64 antennas, respectively.}
%\label{fig:FD-MIMO-Tput}
%\end{figure}

\begin{table}
    \caption{\small FD-MIMO system-level downlink simulation results at 2.5 GHz. Half-wavelength antenna spacings in both the horizontal and vertical dimensions at the BSs, 2 antennas per user, $30\%$ overhead.
%Cell average capacity is the aggregated throughput of the cell and the cell edge throughput is the available throughput for the 5 percentile user.
The baseline is SU-MIMO with 4 antennas per BS and the FD-MIMO results (average and edge data rates) are for MU-MIMO with 16 and 64 antennas, respectively corresponding to $4 \times 4$ and $8 \times 8$ planar arrays per BS sector.}
    \label{tab:FD-MIMO-Tput}
\begin{center}
\begin{tabular}{|l|c|c|c|} \hline
    & SU-MIMO & FD-MIMO 16 & FD-MIMO 64 \\
    \hline\hline
Aggregate Data Rate (b/s/Hz/cell) &  2.32 &  3.28 & 6.37 \\ \hline
Edge Data Rate (b/s/Hz) & 0.063 & 0.1 & 0.4 \\ \hline
\end{tabular}
\end{center}
\end{table}

\subsubsection{Channel Models}

Parallel to the architectural issues run those related to channel
models, which to be sound require extensive field measurements.
Antenna correlations and couplings for massive arrays with relevant
topologies must be determined, and a proper modeling of their impact
must be established; in particular,  the degree of actual channel
orthogonalization in the face of such nonidealities must be
verified. And, for FD-MIMO, besides azimuth, the modeling needs to
incorporate elevation \cite{LU2011,Nam2013,Wu1409:Non}, which is a dimension on
which far less data exists concerning power spectra and angle
spreads. A 3D channel modelling study currently under way within
3GPP is expected to shed light on these various issues
\cite{R1-122034}.
References \cite{Kamm1409:Preliminary,Wu1409:Non} in this special issue also deal with this subject.

\subsubsection{Coexistence with Small Cells}

As mentioned earlier, massive MIMO BSs would most likely have to coexist with tiers of small cells, which would not be equipped with massive MIMO due to their smaller form factor. Although the simplest alternative is to segregate the corresponding transmissions in frequency, the large number of excess antennas at massive MIMO BSs may offer the opportunity of spatial nulling and interference avoidance with relative simplicity and little penalty. 
To confirm the feasibility of this idea, put forth in \cite{ADHIKARY14} and further developed in \cite{Chen1409:Two} within this special issue, comprehensive channel models are again needed. 
% Again,  comprehensive channel models are needed to confirm the feasibility of these approaches.

As networks become dense and more traffic is offloaded to small cells, the number of active users per cell will diminish and the need for massive MIMO may decrease. Aspects such as cost and backhaul will ultimately determine the balance between these complementary ideas.

\subsubsection{Coexistence with mmWave}

As discussed in Sec. \ref{sec-mmWave}, mmWave communication requires many antennas for beamsteering. The antennas are much smaller at these frequencies and thus very large numbers thereof
can conceivably fit into portable devices, and these antennas can indeed provide beamforming power gain but also MIMO opportunities as considered in \cite{Adhi1409:Joint} within this special issue.  Any application of massive MIMO at mmWave frequencies would have to find the correct balance between power gain/interference reduction and parallelization. % \cite{lozano2010transmit}.

\section{Design Issues for 5G }
\label{sec:design}

In addition to supporting 1000x higher throughput, 5G cellular networks must decrease latencies, lower energy consumption, lower costs, and support many low-rate connections.  In this section, we discuss important ongoing research areas that support these requirements.   We begin with the most fundamental aspect of the physical layer---the waveform---and then consider the evolution of cloud-based and virtualized network architectures, latency and control signaling, and energy efficiency.

%=============================
% Waveform subsection (SB)
%=============================
\subsection{The Waveform: Signaling and Multiple Access}

The signaling and multiple access formats, i.e., the waveform design, have changed significantly at each cellular generation and to a large extent they have been each generation's defining technical feature.  They have also often been the subject of fierce intellectual and industrial disputes, which have played out in the wider media. The 1G approach, based on analog frequency modulation with FDMA, transformed into a digital format for 2G and, although it employed both FDMA and TDMA for multiple access, was generally known as ``TDMA'' due to the novelty of time-multiplexing.  Meanwhile, a niche spread spectrum/CDMA standard that was developed by Qualcomm to compete for 2G \cite{GilJac91} became the dominant approach to all global 3G standards.  Once the limitations of CDMA for high-speed data became inescapable, there was a discreet but unmistakable retreat back towards TDMA, with minimal spectrum spreading retained and with the important addition of channel-aware scheduling \cite{BenBla00}. Due to the increasing signal bandwidths needed to support data applications, orthogonal frequency-division multiplexing (OFDM) was unanimously adopted for 4G in conjunction with scheduled FDMA/TDMA as the virtues of orthogonality were viewed with renewed appreciation.

In light of this history, it is natural to ponder the possibility that the transition to 5G could involve yet another major change in the signaling and multiple access formats.
% 5G could adopt yet a different waveform, or perhaps.

%With a renewed appreciation for the cleanness of orthogonal signaling, FDMA/TDMA
%OFDM, which had been discarded for 3G, was then unanimously adopted for 4G.
% (even though later versions of 3G were hardly spread spectrum anymore but rather single carrier TDMA), the consensus for 4G became OFDMA.
%Therefore, it is natural to ponder the possibility that 5G could adopt yet a different waveform.

\subsubsection{OFDM and OFDMA: The Default Approach}

OFDM has become the dominant signaling format for high-speed wireless communication, forming the basis of all current WiFi standards and of LTE, and further of wireline technologies such as digital subscriber lines, digital TV, and commercial radio. Its qualities include:
\begin{itemize}
\item A natural way to cope with frequency selectivity.
\item Computationally efficient implementation via FFT/IFFT blocks and simple frequency-domain equalizers.
\item An excellent pairing for MIMO, since OFDM allows for the spatial interference from multiantenna transmission to be dealt with at a subcarrier level, without the added complication of intersymbol interference.
\end{itemize}
% provides a natural way to cope with frequency selectivity and its implementation is very computationally efficient using FFT/IFFT blocks and simple frequency-domain equalizer structures. It is also an excellent base for MIMO as it allows the spatial interference inherent to multiantenna transmission to be dealt with at a subcarrier level, without the added complication of intersymbol interference.

From a multiple access vantage point, OFDM invites dynamic fine-grained resource allocation schemes in the digital domain, and the term OFDMA is employed to denote orthogonal multiple access at a subcarrier level.
In combination with TDMA, this parcels the time-frequency grid into small units known as resource blocks that can be easily discriminated through digital filtering \cite{OFDMABook}.
%  thus easily allowing a wide range of rates to be supported without waste.
Being able to do frequency and time slot allocation digitally also enables more adaptive and sophisticated interference management techniques such as fractional frequency reuse or spectrum partitions between small cells and macrocells.  Finally, given its near-universal adoption, industry has by now a great deal of experience with its implementation, and tricky aspects of OFDM such as frequency offset correction and synchronization have been essentially conquered.

\subsubsection{Drawbacks of OFDM}
Given this impressive list of qualities, and the large amount of inertia in its favor, OFDM is the unquestionable frontrunner for 5G. However, some weak points do exist that could possibly become more pronounced in 5G networks.

First, the peak-to-average-power ratio (PAPR) is higher in OFDM than in other formats since the envelope samples are nearly Gaussian due to the summation of uncorrelated inputs in the IFFT. Although a Gaussian signal distribution is capacity-achieving under an average power constraint \cite{Sha48}, in the face of an actual power amplifier
a high PAPR
% (notably, however, we know from Shannon \cite{Sha48} that a Gaussian distribution is optimal from a channel capacity point of view).
sets up an unattractive tradeoff between the linearity of the transmitted signal and the cost of the amplifier. This problem can be largely overcome by precoding the OFDM
signals at the cost of a more involved equalization process at the receiver and
% SC-FDMA, as in the uplink of OFDMA, at the cost of not allowing non-contiguous subcarrier allocations and
a slight power penalty; indeed, this is already being done in the LTE uplink \cite{LTEBook}.

Second, OFDM's spectral efficiency is satisfactory, but could perhaps be further improved upon if the requirements of strict orthogonality were relaxed and if
%$1/T$ spacing between consecutive carriers (with $T$ being the block duration),
the cyclic prefixes (CPs) that prevent interblock interference were smaller or discarded.
The paper \cite{Hong1409:Frequency} in this special issue, instead,  proposes the use of a novel OFDMA-based modulation scheme named  frequency and quadrature amplitude modulation (FQAM), which is shown to improve the downlink throughput for cell-edge users.

 % and border guard bands.

Perhaps the main source of concerns, or at least of open questions, is the applicability of OFDM to mmWave spectrum given the enormous bandwidths therein and the difficulty of developing efficient power amplifiers at those frequencies. For example, a paper in this special issue proposes a single-carrier signaling with null cyclic prefix as an alternative to OFDM at mmWave frequencies \cite{ghoshmillimeter}.

\subsubsection{Potential Alternatives to OFDM}

To address OFDM's weaknesses, we now overview some alternative approaches being actively investigated.  Most of these, however, can be considered incremental departures from OFDM rather than the step-function changes that took place in previous cellular generations.

{\bf Time-frequency packing.} Time-frequency packing \cite{TFpacking} and faster-than-Nyquist signaling \cite{FTN1,FTNprocieee,FTN2} have been recently proposed to circumvent the limitations of strict orthogonality and CP. In contrast to OFDM, where the product of the symbol interval and the subcarrier spacing equals 1, in faster-than-Nyquist signaling products smaller than 1 can be accommodated and spectral efficiency improvements on the order of $25\%$ have been claimed.
%The paper \cite{TFpacking} shows that net improvements on the achievable spectral efficiency in the order of 25$\%$ are possible through a subcarrier spacing in the order of $0.8/T$. Similar arguments can be offered in the time domain via time-packing.

{\bf Nonorthogonal signals.} There is a growing interest in multicarrier formats, such as filterbank multicarrier \cite{farhang2011}, that are natively nonorthogonal and thus do not require prior synchronization of distributed transmitters. A new format termed universal filtered multiCarrier (UFMC) %, for instance,
has been proposed whereby, starting with an OFDM signal, filtering is performed on groups of adjacent subcarriers with the aim of reducing sidelobe levels and intercarrier interference resulting from poor time/frequency synchronization \cite{universalFDM,wunder5gnow}.
%The European Project 5GNOW has recently proposed a new modulation format called universal filtered multi-carrier (UFMC) \cite{universalFDM,wunder5gnow}: starting from OFDM, filtering is performed on groups of adjacent subcarriers with the aim of reducing the spectral side-lobe levels and the inter-carrier interference resulting from poor time and frequency synchronization. In particular, \cite{wunder5gnow} claims that UFMC outperforms OFDM in the presence of reciprocal delays exceeding the cyclic prefix length.

{\bf Filterbank multicarrier.} To address the drawbacks of rectangular time windowing in OFDM, namely the need for large guard bands, \cite{comp2013} shows that the use of filterbank multicarrier permits a robust estimation of very large propagation delays and of arbitrarily high carrier frequency offsets, whereas OFDM would have required a very long CP to attain the same performance levels.

{\bf Generalized frequency division multiplexing.} GFDM is a multicarrier technique that adopts a shortened CP through the tail biting technique and is particularly well suited for noncontiguous frequency bands \cite{GFDM1,GFDM2}, which makes it attractive for spectrum sharing where frequency-domain holes may have to be adaptively filled. %It adopts a shortened cyclic prefix through the tail biting technique, has low out-of-band emissions and its flexibility makes it the natural choice in a cognitive setting, wherein non-contiguous spectrum holes (white spaces) may be adaptively filled.

{\bf Single carrier.} Single-carrier transmission has also been attracting renewed interest, chiefly due to the development of low-complexity nonlinear equalizers implemented in the frequency domain \cite{single2010benvenuto}. This may be of particular interest for mmWave as discussed in this same special issue \cite{ghoshmillimeter}.

\begin{figure}[!t]
\centering
\includegraphics[width=0.9777\columnwidth]{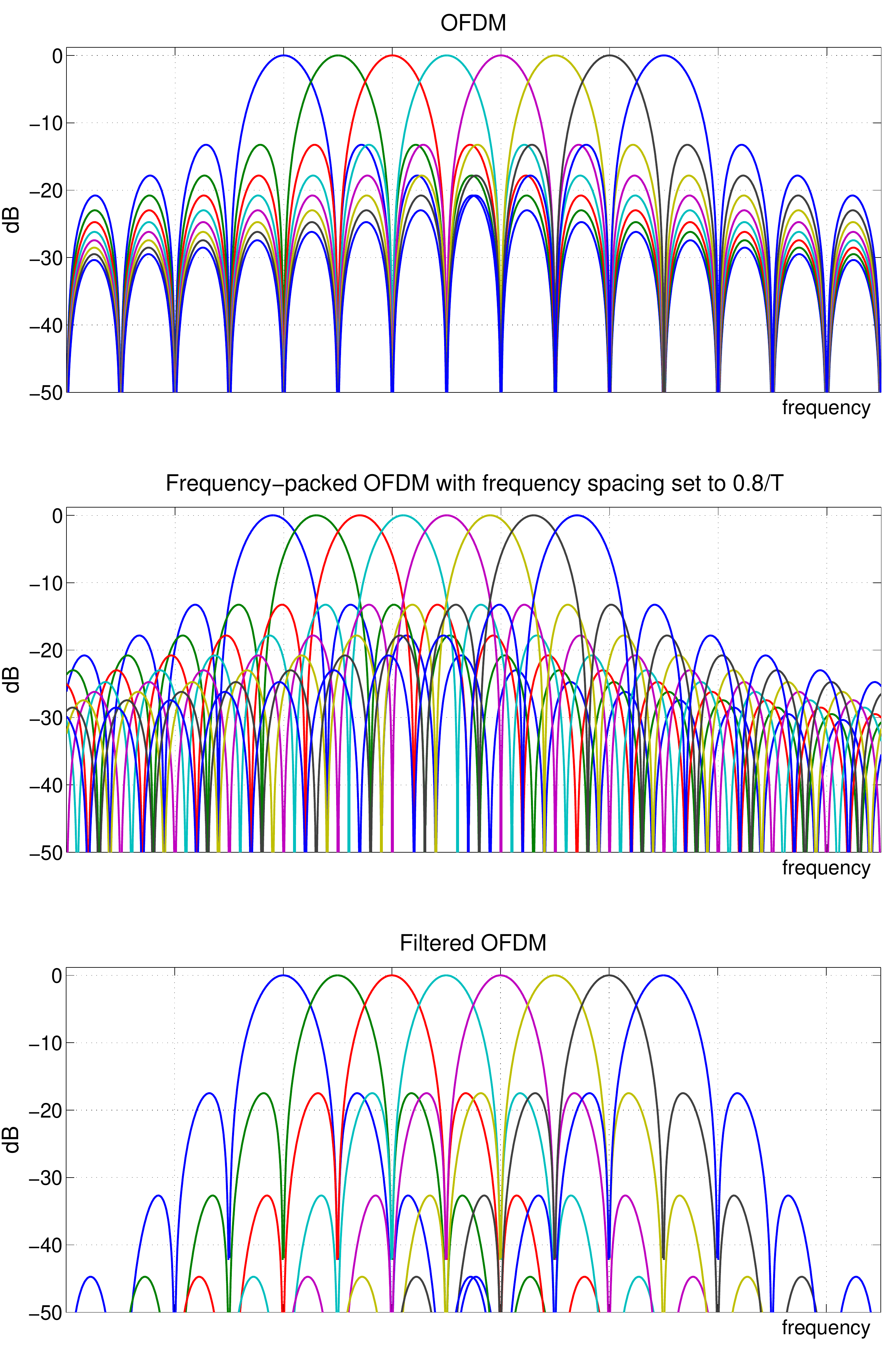}
\caption{Frequency-domain magnitude responses of some adjacent
waveforms for OFDM, frequency-packed OFDM, and filtered OFDM. The
two signaling formats alternative to OFDM trade subcarrier
orthogonality for either better spectral efficiency
(frequency-packed OFDM) or lower out-of-band emissions (filtered
OFDM).} \label{fig:waveform}
\end{figure}

{\bf Tunable OFDM.} We conclude with our own opinion that OFDM could be well adapted to different 5G requirements by %simply
allowing some of its parameters to be %more
tunable, rather than designed for essentially the worst-case multipath delay spread.  In particular, given the increasingly software-defined nature of radios, the FFT block size, the subcarrier spacing and the CP length could change with the channel conditions: in scenarios with small delay spreads---notably dense urban/small cells and mmWave channels---the subcarrier spacing could grow and the FFT size and the CP could be significantly shortened to lower the latency, the PAPR,  the CP's power and bandwidth penalty, and the computational complexity; in channels with longer delay spreads, that could revert to narrower subcarriers, longer FFT blocks, and a longer CP.

%In summary, even though it is possible that OFDM will be still the preferred modulation format for 5G networks, there are non-negligible arguments against such a conservative choice, and chances are that we are just at the beginning of a new physical layer war.

%=============================
% Cloud/Virtualization subsection (SB)
%=============================
\subsection{Cloud-based Networking}
\label{sec:CRAN}

Although this special issue is mainly focused on the air interface, %there is no question that the wide-spread acceptance of smart phones in the last few years is currently transforming the mobile communication network. Consequently,
for the sake of completeness we briefly touch on the exciting changes taking place at the network level.
In that respect, the most relevant event is the movement of data to the cloud so that it can be accessed from anywhere and via a variety of platforms. This fundamentally redefines the endpoints and the time frame for which network services are provisioned. It requires that the network be much more nimble, flexible and scalable. As such, two technology trends will become paramount in the future: network function virtualization (NFV) and software defined networking (SDN). Together, these trends represent the biggest advance in mobile communication networking in the last 20 years, bound to fundamentally change the way network services are provided.

Although the move towards virtualization is thus far taking place only within the core network, this trend might eventually expand towards the edges. In fact, the term cloud-RAN is already being utilized, but for now largely to refer to schemes whereby multiple BSs are allowed to cooperate \cite{CloudRAN}. If and when the Bss themselves become virtualized---down to the MAc and PHY----this term will be thoroughly justified \cite{zhu2011virtual}.  

\subsubsection{Network Function Virtualization}

NFV enables network functions that were traditionally tied to hardware appliances to run on cloud computing infrastructure in a data center. It should be noted that this does not imply that the NFV infrastructure will be equivalent to commercial cloud or enterprise cloud. What is expected is that there will be a high degree of reuse of what commercial cloud offers.

It is natural to expect that some requirements of mobile networks such as the separation of the data plane, control plane and management plane, will not be feasible within the commercial cloud. Nevertheless, the separation of the network functions from the hardware infrastructure will be the cornerstone of future architectures. The key benefit will be the ability to elastically support network functional demands. Furthermore, this new architecture will allow for significant nimbleness through the creation of virtual networks and of new types of network services \cite{NFV12}. A detailed description of the NFV architecture is beyond the scope of this paper, and interested readers can consult \cite{NFV12,NFV13,ATT13} and the references therein.

As virtualization of the communication network gains traction in the industry, an old concept, dating back to the 1990s, will emerge: the provision of user-controlled management in network elements. %, has recently also received significant attention.
Advances in computing technology have reached a level where this vision can become a reality, with the ensuring architecture having recently been termed software defined networking (SDN).

\subsubsection{Software Defined Networking}

SDN is an architectural framework for creating intelligent programmable networks. Specifically, it is defined as an architecture where the control and data planes are decoupled, network intelligence and state are logically centralized, and the underlying network infrastructure is abstracted from the application \cite{ONF12}.

The key ingredients of SDN are an open interface between the entities in the control and data planes, as well as programmability of the network entities by external applications. The main benefits of this architecture are the logical decoupling of the network intelligence to separate software-based controllers, exposing the network capabilities through an application program interface, and enabling the application to request and manipulate services provided by the network \cite{Sezer13}.

From a wireless core network point of view, NFV and SDN should be viewed as tools %that will enable
for provisioning the next generation of core networks %Much research is still needed to determine the overall design.
with many issues still open in terms of scalability, migration from current structures, management and automation, and security.

\subsection{ Energy efficiency}
\label{sec:energy}

%This section discusses the need to keep energy consumption close to what it currently is despite going to 1000x the data.  Very challenging.  I look to SB and others to define the first cut of what should go in this section.  I imagine it will include things like energy harvesting (but let's not overdo it), aggressive sleep modes, new energy-efficient A/D and RF design paradigms for mmWave (huge bandwidths and lots of antennas), what else?
%
%Previous comments:
%\begin{itemize}
%\item AL: Any technique that increases spectral efficiency also increases energy efficiency, as the two can be traded off. It's a matter of selecting an operating point.
%\item JA: leaving this here for now, would like to see what is envisioned for this section though by SB and others.
%\item SB: will take first crack here, and see what happens....
%\end{itemize}
%As it is well known, the topic of energy efficient communications, which, excluded some notable exceptions \cite{verdu}, had been almost completely neglected in the research literature, has recently gained big momentum and a number of special issues, conferences and research projects have been devoted to \emph{green communications} in the last few years -- see, for instance, \cite{greentouch,JSACspecialissue,CMspecialissue}, as the top of the iceberg.

As specified in our stated requirements for 5G, the energy efficiency of the communication chain---typically measured in either Joules/bit or bits/Joule---will need to improve by about the same amount as the data rate just to maintain the power consumption. And by more if such consumption is to be reduces.  This implies a several-order-of-magnitude increase in energy efficiency, which is extremely challenging.   Unsurprisingly, in recent years there has been a surge of interest in the topic of energy efficient communications, as can be seen from the number of recent special issues, conferences and research projects devoted to ``green communications'' \cite{greentouch,JSACspecialissue,CMspecialissue}.   In addition to laudable environmental concerns, it is simply not viable from a logistical, cost or battery-technology point of view to continually increase power consumption.

Due to the rapidly increasing network density (cf. Sect. \ref{sec:offloading}), the access network consumes the largest share of the energy \cite{howmuch2011}. %, so the biggest efforts to increase energy efficiency must be concentrated on this part of the network. The research has focused on several concurrent tracks.
Research has focused on the following areas.

\subsubsection{Resource allocation} %at the physical layer, energy efficiency of a communication link has been mathematically defined as the ratio between a measure of the throughput (i.e. the Shannon achievable rate or the packet success rate) - which is typically an increasing function of the SINR - and a measure of the consumed power. The resulting performance metric is measured in bit/J and represents the number of bits successfully delivered to the receiver for each energy unit used for transmission.
The literature is rich in contributions dealing with the design of resource allocation strategies aimed at the optimization of the system energy efficiency \cite{saraydar2002,poor2007,buzzipoor,potential2012,miao2008,ng2012,vent2014}; the common message of these papers is that, by accepting a moderate reduction in the data rates that could otherwise be achieved, large energy savings can be attained. Within this special issue, \cite{He1409:Leakage} introduces an energy-efficient coordinated beamforming design for HetNets.

\subsubsection{Network Planning} Energy-efficient network planning strategies %have been conceived. These
include techniques for minimizing the number of BSs for  a coverage target \cite{lowcarbon} and the design of adaptive BS %switch on / switch off
sleep/wake algorithms for energy savings \cite{switch1,zhou2009green,cellzooming,blossoming}. The underlying philosophy  of these papers is that, since networks have been designed to meet peak-hour traffic, %a lot of
energy can be saved by (partially) switching off BSs when they have no active users or simply very low traffic. Of course, there are different degrees of hibernation available for a BS\footnote{As an example, a BS serving few users may choose to operate on a reduced set of subcarriers, or it may switch off some of its sectors.} and attention must be paid in order to avoid unpleasant coverage holes; this is usually accomplished through an increase of the transmitted power from nearby BSs.

%Spectrum sharing among different network operators \cite{jorswieck2011resource} is another option for saving energy in low-traffic conditions, while, recently, a separation between signaling BS and data BS has been proposed \cite{greenandsoft2014}: using an always-on macro BS for the signaling, and small micro BSs for delivering the actual data, micro BSs with no users to serve can be switched off, and only switched back on again when a new user request arrives, with no lack of signaling coverage.

\subsubsection{Renewable energy}  Another intriguing possibility is that of BSs powered by renewable energy sources such as solar power \cite{marsan2013}. This is of urgent interest in developing countries lacking a reliable and ubiquitous power grid, but it is also intriguing more broadly as it allows ``drop and play'' small cell deployment (if wireless backhaul is available) rather than ``plug and play''.   A recent paper showed that in a dense HetNet, plausible per-BS traffic loads can actually be served solely by energy harvesting BSs \cite{DhiLi14}.  A more relaxed scenario is considered in \cite{harvest}, where the resource allocation makes efficient use of both renewable and traditional energy sources.

\subsubsection{Hardware solutions} Finally, much of the power consumption issues will be dealt with by hardware engineers, %including 
with recent work in low-loss antennas, antenna muting, and adaptive sectorization according to traffic requirements %, have been investigated 
(see, e.g., \cite{muting}). \\   %Big energy savings can be obtained here; these aspects are however out of the scope of this paper and we do not discuss them any further.\\

%Despite such huge bulk of activities,
In summary, energy efficiency will be a major research theme for 5G, spanning many of the other topics in this article:
\begin{itemize}
\item  True cloud-RAN could provide an additional opportunity for energy efficiency since the centralization of the baseband processing might save energy \cite{greenandsoft2014}, especially if advances on green data centers are leveraged \cite{greendatacenter}.
\item The tradeoff between having many small cells or fewer macrocells given their very different power consumptions is also of considerable interest \cite{LiZha14}.  
\item A complete characterization of the energy consumed by the circuitry needed for massive MIMO is %also 
currently lacking. %Finally, 
\item MmWave energy efficiency will be particularly crucial given the unprecedented bandwidths \cite{RanCTW13}.
\end{itemize}

%********************************************************************************
% Spectrum and Regulatory and Standards Section (SH, AS, CA)
%********************************************************************************
%\section{Spectrum and Regulatory Innovation (3 pages)}
%\label{sec-spectrum_regulation}

\section{Spectrum, Regulation and Standardization for 5G}

%As encapsulated in Shannon's famous formula for the capacity of the additive, white, Gaussian noise channel, wireless capacity is delivered from two basic resources: radio spectrum, and received signal power.
%To meet the growing demands of data-hungry wireless devices, 5G networks will need to use {\it more} spectrum {\it more} efficiently than today. Spectrum regulation is a key component in the design of future networks.

Departing from strictly technical issues, we now turn our attention to the crucial intersections that 5G technologies will encounter with public policy, industry standardization, and economic considerations.

\subsection{Spectrum Policy and Allocation}
\label{sec-spectrum}
\label{sec-allocation}

As discussed in Section~\ref{sec-mmWave}, the beachfront microwave spectrum is already saturated in peak markets at peak times while large amounts of idle spectrum do exist in the mmWave realm. Due to the different propagation characteristics, and recalling the concept of phantom cells, future systems will need to integrate a broad range of frequencies: low frequencies for wide coverage, mobility support, and control, and high frequencies for small cells.
This will require new approaches to spectrum policy and allocation methods. Topics such as massive MIMO and small cells, which address the efficient use of spectrum, must also be considered important issues in spectrum policy.  Needless to say, spectrum allocation and policy is an essential topic for 5G, so this section considers the pros and cons of different approaches to spectrum regulation in that context. % that may be better suited to future scenarios.

\subsubsection{Exclusive Licenses}

The traditional approach to spectrum policy is for the regulator to award an exclusive license to a particular band for a particular purpose, subject to limitations (e.g., power levels or geographic coverage). Exclusive access gives full interference management control to the licensee and provides an incentive for investments in infrastructure, allowing for quality-of-service guarantees.  Downsides include high entry barriers %are high
because of elevated sunk costs, both in the spectrum itself and in infrastructure, and that such allocations are inherently inefficient since they occur over very long time scales---typically decades---and thus the spectrum is rarely allocated to the party able to make the best economic use of it.
%Typically, the regulator will charge a fee, which can be very high if the licensee can re-use the spectrum over a wide geographic area.

%The advantage of an exclusive license is that interference can be managed by the operator, who is motivated to invest in infrastructure such as base stations and backhaul to provide much needed capacity.
%With an exclusive license, interference can be controlled and there is an incentive for the operator to optimize the network. Quality of service can be guaranteed.

%This is argued to be a problem today, with the huge growth in demand for mobile broadband, but relatively limited available spectrum. Much of the useful spectrum was allocated long ago to other applications ({\it eg.} TV and radio broadcast). In some cases, the bands are occupied by legacy equipment operating with low spectral efficiency. Exclusively allocated spectrum can also be under-utilized.

To address these inefficiencies, market-based approaches have been propounded  \cite{Coase59}. Attempting to implement this idea, spectrum auctions have been conducted recently to refarm spectrum, a process whereby long-held commercial radio and TV allocations are moved to different (smaller) bands releasing precious spectrum for wireless communications; a prime example of this is the so-called ``digital dividend'' auctions arising from the digitization of radio and TV. However, there are claims that spectrum markets have thus far not been successful in providing efficient allocations because such markets are not sufficiently fluid due to the high cost of the infrastructure \cite{Benkler12}. According to these claims, spectrum and infrastructure cannot be easily decoupled.

%However, whilst the allocation may be reasonably efficient at the time of auction, this will fail over the long time intervals between such re-allocations. Further, the markets are not fluid due to the associated infrastructure being very expensive. In reality, spectrum and infrastructure cannot easily be decoupled \cite{Benkler12}.
%
%Also noteworthy in the licensed arena is that regulators have recently been refarming spectrum, moving long-held commercial radio and TV allocations to new bands (typically much smaller) and releasing precious spectrum for wireless communications; often, this released spectrum is auctioned to the highest bidder. A prime example of this is the so-called ``digital dividend'' auctions arising from the digitization of radio and TV.

\subsubsection{Unlicensed Spectrum}

%The ``open access'' approach may not offer an incentive for investment in network infrastructure, because the benefits of a smart network are shared by other parties, and interference management is much more difficult when many distributed entities share the same band.

At the other extreme, regulators can designate a band to be ``open access'', meaning that there is no spectrum license and thus users can share the band provided their devices are certified (by class licenses). Examples are the industrial, scientific and medical (ISM) bands, which are utilized by many devices including microwave ovens, medical devices, sensor networks, cordless phones and especially by WiFi.  With open access, barriers to entry are much lower and there is enhanced competition and innovation, as the incredible success of WiFi and other ISM-band applications makes plain.

The downside of open access is potentially unmanageable interference, no quality-of-service guarantees, and, possibly, the ``tragedy of the commons," where no one achieves a desired outcome.  Still, it is useful to consider the possibility of open access for bands utilized in small cells as future networks may involve multiple players and lower entry barriers may be needed to secure the emergence of small-cell infrastructures.

Although interference is indeed a significant problem in current open access networks, it is interesting to note that cellular operators nevertheless rely heavily on WiFi offloading: currently about half of all cellular data traffic is proactively offloaded through unlicensed spectrum \cite{CiscoVNI14}. WiFi hotspots are nothing but small cells that spatially reuse ISM frequencies. At mmWave frequencies, the main issue is signal strength rather than interference, and it is therefore plausible that mmWave bands be unlicensed, or at a minimum several licensees will share a given band under certain new regulations.  This question is of pressing interest for 5G.

\subsubsection{Spectrum Sharing}
\label{messi}
%
%Options do exist halfway between exclusive licenses and open access. One such option is TV white space, in which the license holder and primary user is a TV station, but secondary users are allowed in the same spectrum provided they do not cause significant interference to the primary service \cite{FCC2010}. This is typically accomplished by means of a data base indicating the primary allocations. Secondary users must gain approval from the regulator, via that data base, to utilize the spectrum at a particular time and geographical location. In essence then, a white space is where spectrum is under-utilized. Not all TV channels broadcast 24/7 everywhere, and hence secondary use can be permitted at certain times and places.
%The white space approach can be extended to other primary services such as radar, meteorology, or naval defense (only needed near the coasts).
%
Options do exist halfway between exclusive licenses and open access, such as the opportunistic use of TV white space. While the potential of reusing this spectrum is enticing, it is not crystal clear that reliable communication services can be delivered that way. Alternatively, Authorized Shared Access~\cite{VTM_ASA13} and Licensed Shared Access~\cite{LSA_12} are regulatory frameworks that allow spectrum sharing by a limited number of parties each having a license under carefully specified conditions. %There is an incumbent primary user but, different from TV whitespace, each ASA licensee has exclusive access rights when and where the incumbent is not active.
Users agree on how the spectrum is to be shared, seeking interference protection from each other, %as well as from the incumbent
thereby increasing the predictability and reliability of their services. %In addition, ASA may facilitate harmonization of spectrum across countries and efficient spectral reuse \cite{QualcommASA13}.

%ASA enables a 5G operators to get more spectrum, and it will provide a better harmonization of spectrum across different countries. The spectrum can be shared by allowing activity at different times, in different frequency bands, or by taking advantage of geographical separation. It includes an ASA database, to record the actual spectrum usage, and an ASA controller to compute spectrum availability, based on the agreed rules and the information from the database.
%It may be particularly useful in allowing spectrum to be re-used more efficiently, where a service provider gets an ASA license to use a band in its small cells, for example. This approach can be enhanced by carrier aggregation, providing substantial increases in data rates for small cell users \cite{QualcommASA13}.

\subsubsection{Market-Based Approaches to Spectrum Allocation}

Given the advantages of exclusive licenses for ensuring quality of service, it is likely that most beachfront spectrum will continue to be allocated that way.
%This spectrum will be used for macro-cell coverage, for control signals, and for highly mobile traffic.
Nevertheless, better utilization could likely be obtained if spectrum markets could become more fluid \cite{Coase59}.
To that end, liberal licenses do not, in principle, preclude trading and reallocation on a fast time scale, rendering spectrum allocations much more dynamic. Close attention must be paid to the definition of spectrum assets, which have a space as well as a time scale, and the smaller the scales, the more fluid the market \cite{ZBHV_JSAC13}.

In small cells, traffic is much more volatile than in macrocells and operators may find it beneficial to enter into sharing arrangements for both spectrum and infrastructure. Dynamic spectrum markets may emerge, managed by brokers, allowing licenses to spectrum assets to be bought and sold---or leased---on time scales of hours, minutes or even ms \cite{BHVCommsMag10}. Along these lines, an interesting possibility is for a decoupling of infrastructure, spectrum and services \cite{BHVCommsMag10}. In particular, there may be a separation between spectrum owners and operators. Various entities may own and/or share a network of BSs, and buy and sell spectrum assets from spectrum owners, via brokers. These network owners may offer capacity to operators, which in turn would serve the end customers with performance guarantees. All of this, however, would require very adaptable and frequency agile radios.
%A smart phone today contains several different radios (for different 3G and 4G networks, for WiFi, and GPS), and 5G radios will offer a much more complete integration of many different technologies, including the possibility of mmWave communications. Frequency-agility is likely to be an important characteristic of a 5G device.

We conclude this discussion by noting that offloading onto unlicensed spectrum such as TV whitespace or mmWave bands could have unexpected results.  In particular, adding an unlicensed shared band to an environment where a set of operators have exclusive bands can lead to an overall \emph{decrease} in the total welfare (Braess' paradox) \cite{NZBHV11}. This is because operators might have an incentive to offload traffic even when this runs counter to the overall social welfare, defined as the total profit of the operators and the utilities of the users, minus the costs. An operator might have an incentive to increase prices so that some traffic is diverted to the unlicensed band, where the cost of interference is shared with other operators, and this price increase more than offsets the operator's benefits. Further,
while unlicensed spectrum generally lowers barriers to entry and increases competition, the opposite could occur and in some circumstances a single monopoly operator could emerge \cite{ZBHV12} within the unlicensed bands.

\subsection{Regulation and Standardization}
\label{sec:standards}

\subsubsection{5G Standardization Status}

%5G has gained a lot of attention in the last couple years, and
Several regional forums and projects have been established to shape the 5G vision and to study its key enabling technologies~\cite{METIS,ChinaPromotion,Arib,5GKorea}.  For example, the aforementioned EU project METIS has already released documents on scenarios and requirements~\cite{METIS12013,METIS22013}. Meanwhile,  5G has been increasingly referred to as ``IMT-2020'' in many industry forums and international telecommunications union (ITU) working groups \cite{ITU-R2014-1} with the goal, as the name suggests, of beginning commercial deployments around 2020.
%To achieve this goal,  ITU-R working party 5D has set a target to complete a vision document as a draft new recommendation to ITU-R by June 2015. This vision document, titled ``Framework and overall objectives of the future development of IMT for 2020 and beyond'', is to help generate momentum and accelerate global research, technology development, and standardization activities, with a goal of successful launch of 5G systems around 2020.

To explore 5G user requirements and to elaborate a standards agenda to be driven by them, the ETSI held a future mobile summit~\cite{Scrase} in Nov. 2013. The summit concluded, in line with the thesis of this paper, that an evolution of LTE may not be sufficient to meet the anticipated 5G requirements. That conclusion notwithstanding, 5G standardization has not yet formally started within 3GPP, which is currently finalizing LTE Rel-12 (the third release for the LTE-Advanced family of 4G standards). The timing of 5G standardization has not even been agreed upon, although it is not expected to start until later Rel-14 or Rel-15, likely around 2016--2017. However, many ongoing and proposed study items for Rel-12 are already closely related to 5G candidate technologies covered in this paper (e.g., massive MIMO) and thus, in that sense, the seeds of 5G are being planted in 3GPP.  Whether an entirely new standards body will emerge for 5G as envisioned in this paper is unclear; the ongoing success of 3GPP relative to its erstwhile competitors (3GPP2 and the WiMAX Forum) certainly gives it an advantage, although a name change to 5GPP would seem to be a minimal step.

\subsubsection{5G Spectrum Standardization}

Spectrum standardization and harmonization efforts for 5G have begun within the ITU.
% ITU working party 5D (WP5D~\cite{ITU-R2014-2})
Studies are under way on the feasibility of  bands above 6~GHz~\cite{ITU-R2014-2}, including technical aspects such as channel modelling, semiconductor readiness, coverage, mobility support, potential deployment scenarios and coexistence with existing networks.
%The motivation is straightforward: if we can make use of %these
%high frequency spectrum that is relatively abundant, %at higher frequencies,
%then much of  the spectrum shortage we are experiencing today can be alleviated (elaborated in Section~\ref{sec-allocation}). This will encourage continued innovation and investment in the wireless space,  and further improve the mobile computing experience for billions of users around the world. %in Section~\ref{sec-spectrum}.

%The semiconductor industry has made major progress on mmWave technology  in the last decade, and mmWave wireless solutions and products are now available in the 60GHz band for high-rate, short distance communications (known as WiGiG devices),  in the 28 and 38 GHz for fixed wireless communications,  and  in the E-band (71-76, 81-86 GHz) for wireless backhauling.  GaAs MMIC (Monolithic Microwave Integrated Circuit) is improving fast and is expected to be mature enough in the next few years to provide commercial-grade power-efficient amplifiers, LNAs, VCOs, phase shifters and other radio frequency integrated circuit solutions.

To be available for 5G, mmWave spectrum has to be repurposed by national regulators for mobile applications and agreement must be reached in ITU world radiocommunication conferences (WRC) on the global bands %therefore.
for mmWave communications. These processes tend to be tedious and lengthy, and there are many hurdles to clear before the spectrum can indeed be available.
On the ITU side,
%it is all but certain that in the upcoming WRC-15 meeting the focus of discussion will be on below $6$~GHz bands, as a first step to relieve the critical shortage of spectrum for the industry.
WRC-18 is shaping up as the time and venue to agree on mmWave spectrum allocations for 5G.
%In fact, recently several proposals  were submitted to EU regional WRC preparation meetings, with the goal of getting higher frequency spectrum for 5G agreed in WRC-18.

In addition to the ITU, many national regulators have also started their own studies on mmWave spectrum for mobile communications. In the USA, the technological advisory council of the federal communications committee (FCC)
has carried out extensive investigations on mmWave technology in the last few years and it is possible that FCC will issue a notice of inquiry in 2014, which is always the first step in FCC's rulemaking process for allocation of any new frequency bands.  As discussed above, it is also unclear how such bands will be allocated or even how they \emph{should} be allocated, and the technical community should actively engage the FCC to make sure they are allocated in a manner conducive to meeting 5G requirements.  Historically, other national regulators have tended to follow the FCC's lead on spectrum policy.

%\subsection{Other Regulatory Issues }
%\label{sec-regulation}

\subsection{Economic Considerations}
\label{sec-economics}

The economic costs involved in moving to 5G are substantial.   Even if spectrum costs can be greatly reduced through the approaches discussed above, it is still a major challenge for carriers to densify their networks to the extent needed to meet our stated 5G requirements.  Two major challenges are that BS sites are currently expensive to rent, and so is the backhaul needed to connect them to the core network.

\subsubsection{Infrastructure Sharing}

One possible new business model could be based on infrastructure sharing, where the owners of infrastructure and the operators are different. There are several ways in which infrastructure could be shared.

\textbf{Passive sharing.} The passive elements of a network include the sites (physical space, rooftops, towers, masts and pylons), the backhaul connection, power supplies, and air-conditioning. Operators could cover larger geographical areas at a lower cost and with less power consumption if they shared sites, and this might be of particular importance in dense 5G networks \cite{MRG11}. Regulation could be required to force major operators to share their sites and improve competition.

\textbf{Active sharing.} Active infrastructure sharing would involve antennas, BSs, radio access networks and even core networks.
BS and/or radio access network sharing may be particularly attractive when rolling out small-cell networks \cite{GMW13}.  This type of sharing could lead to collusion, with anticompetitive agreements on prices and services \cite{MRG11}. Regulations are required to prevent such collusion, but on the positive side are the economies of scale.

\textbf{Mobile virtual network operators.}  A small cell may be operated by a mobile virtual network operator that does not own any spectrum
but has entered into an agreement with another operator to gain access to its spectrum within the small cell. The small cell may provide coverage to an enterprise or business such that, when a user leaves the enterprise, it roams onto the other operator's network.

\textbf{Offloading.} Roaming is traditionally used to increase coverage in scenarios when service providers' geographical reaches are limited. However, in 5G, and as discussed above, traffic may be offloaded for a different reason: spatial and temporal demand fluctuations. Such fluctuations will be greater in small-cell networks. Recent papers consider the incentive for investment under various revenue-sharing contracts \cite{BHNSZV13,LMK14}. It is shown in \cite{BHNSZV13} that sharing increases investment, and the incentive is greater if the owner of the infrastructure gets the larger fraction of the revenue when overflow traffic is carried. A bargaining approach for data offloading from a cellular network onto a collection of WiFi or femtocell networks is considered in \cite{Gao1409:Bargaining} in this special issue.

\subsubsection{Backhaul}  A major consideration that has been considered in several places throughout the paper is backhaul, which will be more challenging to provide for hyper-dense ultra-fast networks.  However, we find optimism in three directions.
\begin{itemize}
\item Fiber deployments worldwide continue to mature and reach farther and farther into urban corridors.
\item Wireless backhaul solutions are improving by leaps and bounds, with considerable startup activity driving innovation and competition.  Further, mmWave frequencies %themselves
could be utilized for much of the small-cell backhauling due to their ambivalence to interference.  This may in fact be the first serious deployment of non-LoS mmWave with massive beamforming gains given that the backhaul connection is quite static and outdoors-to-outdoors, and thus more amenable to precise beam alignment.
\item Backhaul optimization is becoming a pressing concern, given its new status as a performance-limiting factor, and this is addressed in \cite{Liao1409:Min,Zhou1409:Optimized} in this special issue. The problem of jointly optimizing resources in the radio network and across the backhaul is considered in \cite{Liao1409:Min}.  Compression techniques for uplink cloud-RAN are developed in \cite{Zhou1409:Optimized}. Another approach is the proactive caching of high bandwidth content like popular video \cite{GolSha11}.
\end{itemize}

\section{Conclusions}

It is an exciting time in the wireless industry and for wireless research at large.  Daunting new requirements for 5G are already unleashing a flurry of creative thinking and a sense of urgency in bringing innovative new technologies into reality.  Even just two years ago, a mmWave cellular system was considered something of a fantasy; now it is almost considered an inevitability.  As this article has highlighted, it is a long road ahead to truly disruptive 5G networks.  Many technical challenges remain spanning all layers of the protocol stack and their implementation, as well as many intersections with regulatory, policy, and business considerations.  We hope that this article and those in this special issue %in time will be seen to have
will help to move us forward along this road.

%\section*{Acknowledgments}
%
%
%Thank authors, reviewers for the special issue.  Also our colleagues if they contributed tangentially, etc...
%
%For page counts, assume 1 page for combo of title + abstract, conclusions, and Acks.   3 pages of Refs.

\section*{Acknowledgments}

The authors thank Arunabha Ghosh (AT\&T Labs), Robert W. Heath Jr. (UT Austin), and Federico Boccardi (Vodaphone) for very helpful feedback and suggestions on the paper.

\bibliographystyle{IEEEtran}
\bibliography{Bib/Andrews,Bib/Choi,Bib/Hanly,Bib/Charlie-Anthony,Bib/buzzi,Bib/AL}

\begin{IEEEbiography}[{\includegraphics[width=1in,height=1.25in,clip,keepaspectratio]{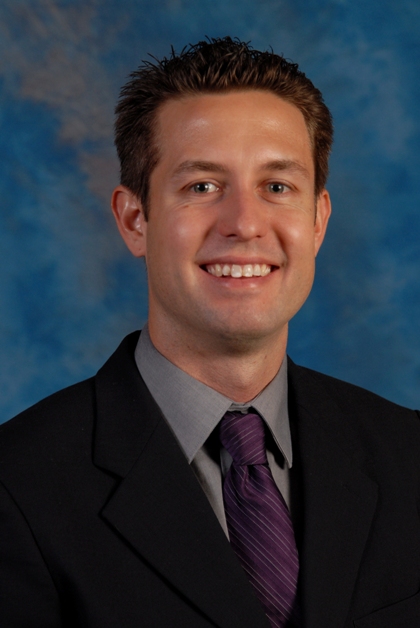}}]
%[{\includegraphics[width=1in,height=1.25in,keepaspectratio]{Pics/.jpg}]
{Jeffrey~G.~Andrews} [S'98, M'02, SM'06, F'13] received the B.S. in
Engineering with High Distinction from Harvey Mudd College, and the
M.S. and Ph.D. in Electrical Engineering from Stanford University.
He is the Cullen Trust Endowed Professor (\#1) of ECE at the
University of Texas at Austin, Editor-in-Chief of the IEEE
Transactions on Wireless Communications, and Technical Program
Co-Chair of IEEE Globecom 2014. He developed CDMA systems at
Qualcomm from 1995-97, and has consulted for entities including
Verizon, the WiMAX Forum, Intel, Microsoft, Apple, Samsung,
Clearwire, Sprint, and NASA.  He is a member of the Technical
Advisory Boards of Accelera and Fastback Networks, and co-author of
the books Fundamentals of WiMAX and Fundamentals of LTE.

Dr. Andrews received the National Science Foundation CAREER award in
2007 and has been co-author of nine best paper award recipients: ICC
2013, Globecom 2006 and 2009, Asilomar 2008, the 2010 IEEE
Communications Society Best Tutorial Paper Award, the 2011 IEEE
Heinrich Hertz Prize, the 2014 EURASIP Best Paper Award, the 2014
IEEE Stephen O. Rice Prize, and the 2014 IEEE Leonard G. Abraham
Prize.  He is an elected member of the Board of Governors of the
IEEE Information Theory Society.

\end{IEEEbiography}
\begin{IEEEbiography} [{\includegraphics[width=1in,height=1.25in,clip,keepaspectratio]{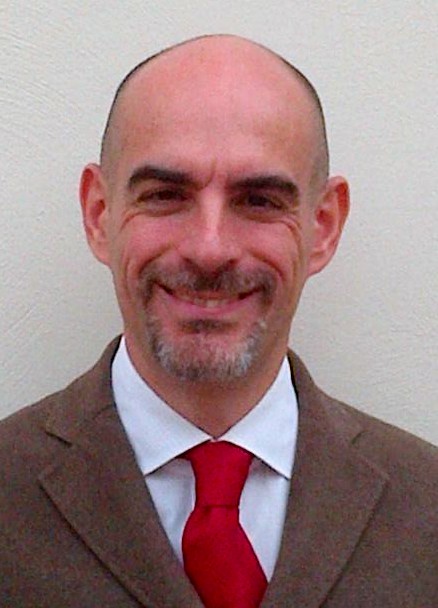}}]
{Stefano~Buzzi} [M'98, SM'07] is currently an Associate Professor at
the University of Cassino and Lazio Meridionale, Italy. He received
his Ph.D. degree in Electronic Engineering and Computer Science from
the University of Naples "Federico II" in 1999, and he has had
short-term visiting appointments at the Dept. of Electrical
Engineering, Princeton University, in 1999, 2000, 2001 and 2006. His
research and study interest lie in the wide area of statistical
signal processing and resource allocation for communications, with
emphasis on wireless communications; he  is author/co-author of more
than 50 journal papers and 90 conference papers; Dr. Buzzi was
awarded by the Associazione Elettrotecnica ed Elettronica Italiana
(AEI) the "G. Oglietti" scholarship in 1996, and was the recipient
of a NATO/CNR advanced fellowship in 1999 and of three CNR
short-term mobility grants. He is a former Associate Editor for the
\emph{IEEE Communications Letters}, and the {\em IEEE Signal
Processing Letters}.
\end{IEEEbiography}
\begin{IEEEbiography}[{\includegraphics[width=1in,height=1.25in,clip,keepaspectratio]{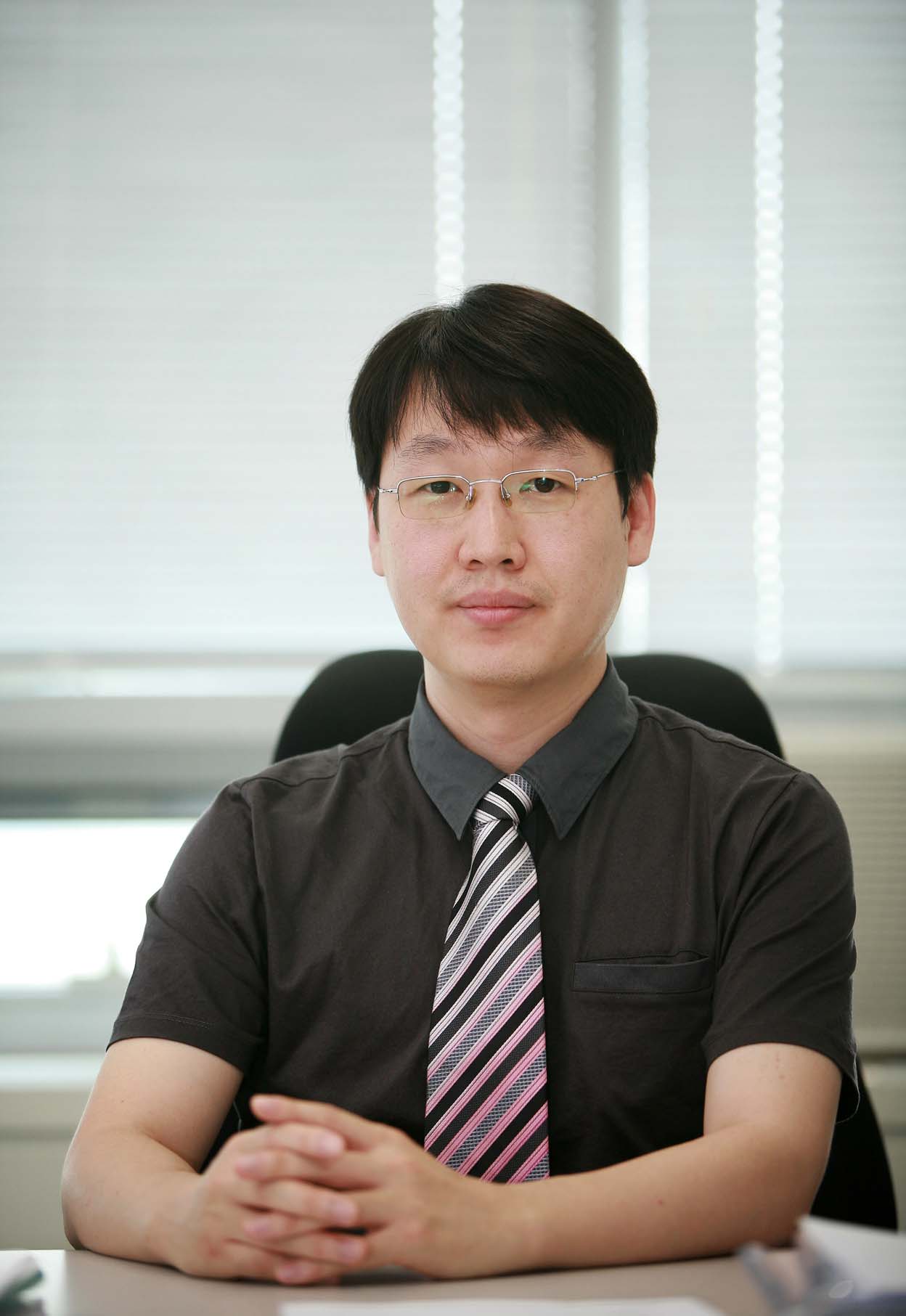}}]
{Wan~Choi} [M'06, SM'12] received the B.Sc. and M.Sc. degrees from
the School of Electrical Engineering and Computer Science (EECS),
Seoul National University (SNU), Seoul, Korea, in 1996 and 1998,
respectively, and the Ph.D. degree in the Department of Electrical
and Computer Engineering at the University of Texas at Austin in
2006. He is currently an Associate Professor of the Department of
Electrical Engineering, Korea Advance Institute of Science and
Technology (KAIST), Daejeon, Korea. From 1998 to 2003, he was a
Senior Member of the Technical Staff of the R\&D Division of KT
Freetel, Korea, where he researched 3G CDMA systems.

Dr. Choi is the recipient of IEEE Vehicular Technology Society Jack
Neubauer Memorial Award in 2002. He also received the IEEE Vehicular
Technology Society Dan Noble Fellowship Award in 2006 and the IEEE
Communication Society Asia Pacific Young Researcher Award in 2007.
He serves as Associate Editor for the IEEE Transactions on Wireless
Communications, for the IEEE Transactions on Vehicular Technology,
and for IEEE Wireless Communications Letters.
\end{IEEEbiography}
\begin{IEEEbiography}[{\includegraphics[width=1in,height=1.25in,clip,keepaspectratio]{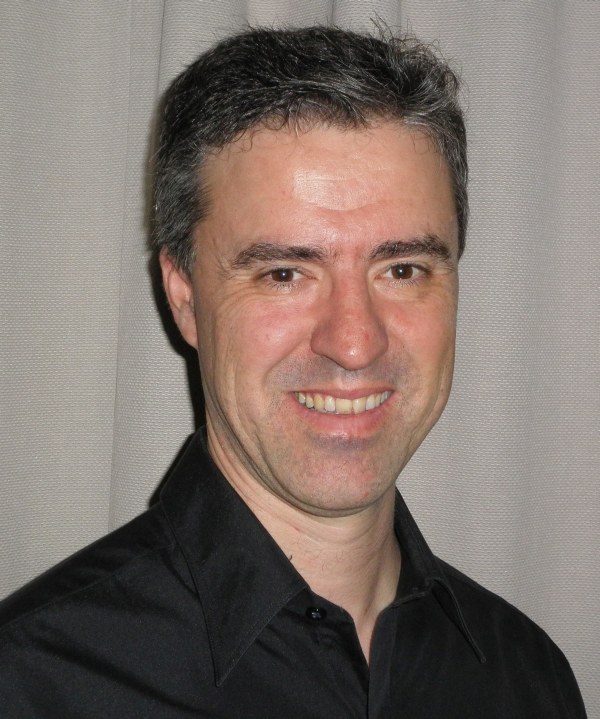}}]
{Stephen~V.~Hanly} [M'98] received a B.Sc. (Hons) and M.Sc. from the
University of Western Australia, and the Ph.D. degree in mathematics
in 1994 from Cambridge University, UK. From 1993 to 1995, he was a
Post-doctoral member of technical staff at AT\&T Bell Laboratories.
From 1996-2009 he was at the University of Melbourne, and from
2010-2011 he was at the National University of Singapore. He now
holds the CSIRO-Macquarie University Chair in Wireless
Communications at Macquarie University, Sydney, Australia. He has
been an Associate Editor for IEEE Transactions on Wireless
Communications, Guest Editor for IEEE Journal on Selected Areas in
Communications, and Guest Editor for the Eurasip Journal on Wireless
Communications and Networking. In 2005 he was the technical co-chair
for the IEEE International Symposium on Information Theory held in
Adelaide, Australia. He was a co-recipient of the best paper award
at the IEEE Infocom 1998 conference, and the 2001 Joint IEEE
Communications Society and IEEE Information Theory Society best
paper award.
\end{IEEEbiography}
\begin{IEEEbiography}[{\includegraphics[width=1in,height=1.25in,clip,keepaspectratio]{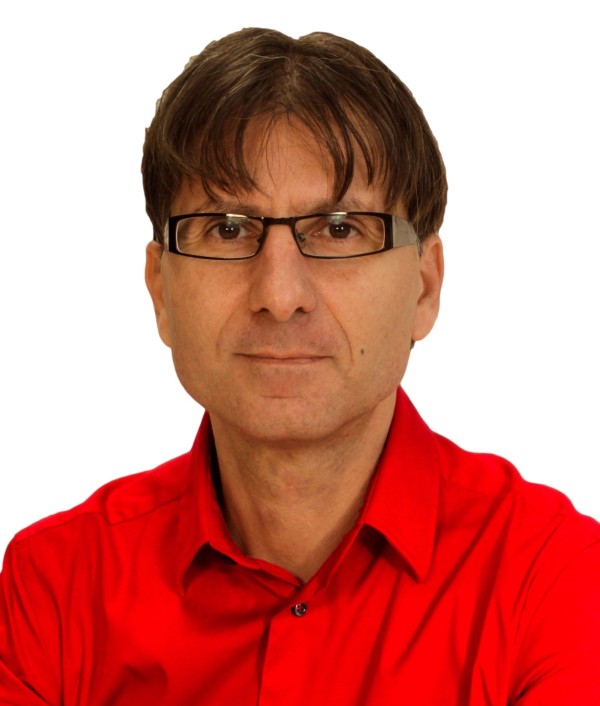}}]
 {Angel~Lozano} [S'90, M'99, SM'01, F'14] received the M.Sc. and Ph.D. degrees in
Electrical Engineering from Stanford University in 1994 and 1998,
respectively. He is a Professor and the Vice-Rector for Research at
Universitat Pompeu Fabra (UPF) in Barcelona, Spain. He was with Bell
Labs (Lucent Technologies, now Alcatel-Lucent) between 1999 and
2008, and served as Adj. Associate Professor at Columbia University
between 2005 and 2008. Prof. Lozano is an Associate Editor for the
IEEE Transactions on Information Theory, the Chair of the IEEE
Communication Theory Technical Committee, and an elected member to
the Board of Governors of the IEEE Communications Society. His
papers have received two awards: ISSSTA 2006 and the 2009 IEEE
Stephen O. Rice prize.
\end{IEEEbiography}
\begin{IEEEbiography}[{\includegraphics[width=1in,height=1.25in,clip,keepaspectratio]{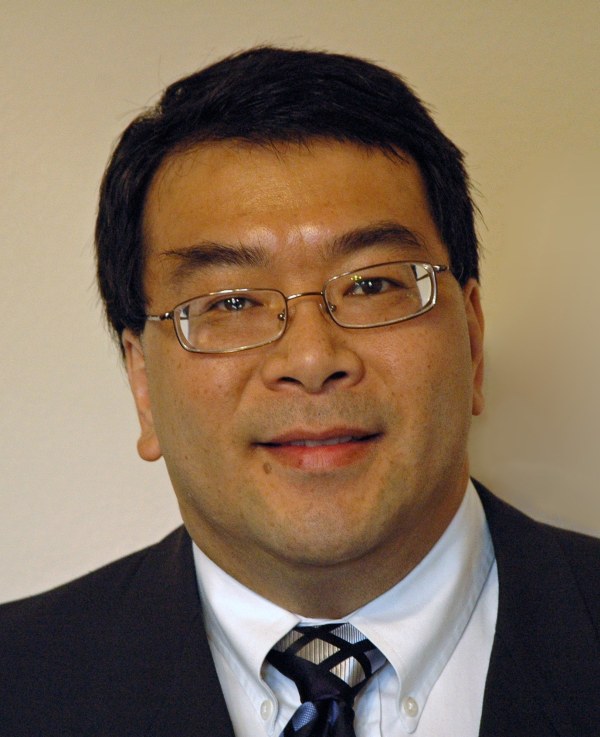}}]
{Anthony C. K. Soong} [S'88, M'91, SM'02, F'14] received the B.Sc. degree in animal physiology and physics from the University of Calgary, and the B.Sc. degree in electrical engineering, the M.Sc. degree in biomedical physics and Ph.D. degree in electrical and computer engineering from the University of Alberta. He is currently the chief scientist for wireless research and standards at Huawei Technologies Co. Ltd, in the US. He serves as the vice-chair for 3GPP2 TSG-C WG3. Prior to joining Huawei, he was with the systems group for Ericsson Inc and Qualcomm Inc. Dr. Soong has published numerous scientific papers and has over 80 patents granted or pending. He was the corecipient of the 2013 IEEE Signal Processing Society Best Paper Award.
\end{IEEEbiography}

\begin{IEEEbiography}[{\includegraphics[width=1in,height=1.25in,clip,keepaspectratio]{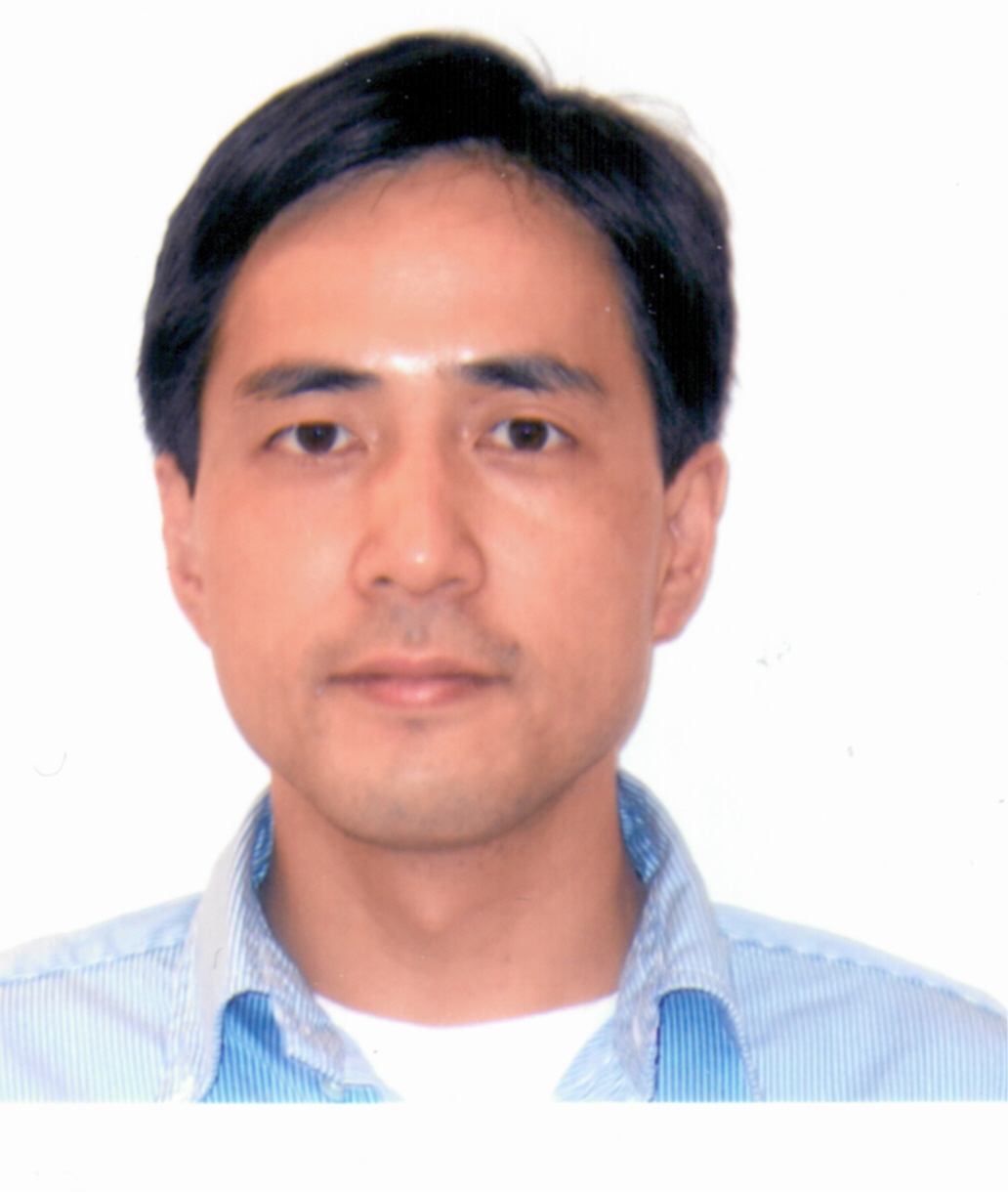}}]
{Jianzhong (Charlie) Zhang} [S'96, M'02, SM'09]  is currently senior director and head of Wireless Communications Lab with Samsung Research America at Dallas, where he leads technology development, prototyping and standardization for Beyond 4G and 5G wireless systems. From Aug 2009 to Aug 2013, he served as the Vice Chairman of the 3GPP RAN1 working group and led development of LTE and LTE-Advanced technologies such as 3D channel modeling, UL-MIMO and CoMP, Carrier Aggregation for TD-LTE, etc. Before joining Samsung, he was with Motorola from 2006 to 2007 working on 3GPP HSPA standards, and with Nokia Research Center from 2001 to 2006 working on IEEE 802.16e (WiMAX) standard and EDGE/CDMA receiver algorithms. He received his Ph.D. degree from University of Wisconsin, Madison.
 \end{IEEEbiography}

\end{document}